\documentclass{raa_twocolumn} 

\usepackage{graphicx,times}             %for PS/EPS graphics inclusion, new
\usepackage{natbib}
\usepackage{amssymb,amsmath}
\usepackage[pagebackref=true]{hyperref}
\usepackage{array}
\usepackage{booktabs} % pour des lignes horizontales élégantes
\usepackage{longtable}
\usepackage{comment} % to have the capacity to add commented zone with \begin{comment}
\usepackage{float}
\usepackage{booktabs}

\begin{document}

  \title{The VHF alert network of the SVOM mission}
%   \subtitle{I. Place Your Subtitle Here}

   \volnopage{Vol.0 (202x) No.0, 000--000}      %%preserved for Editor. DOn't remove!
   \setcounter{page}{1}          %%starting page, preserved for Editor. DOn't remove!

   \author{B. Cordier
      \inst{1,*}\footnotetext{$*$Corresponding Author}
    \and L. Jeannin
      \inst{3}
    \and Ph. Lafabrie
      \inst{3} %\footnotetext{ID: https:/orcid.org/0000-0002-0003-0004}
    \and G. Chavanas
      \inst{3}    
    \and S. Crepaldi
      \inst{3}
    \and N. Dagoneau
      \inst{2}
    \and A. Formica
      \inst{2}
    \and V. Garcia
      \inst{3}
    \and L. Jolivet
      \inst{3}
    \and S. Lacour
      \inst{3}
    \and H. Louvin
      \inst{2}
    \and E. Sabatier
      \inst{3}
   }

   \institute{CEA/Paris-Saclay, Irfu/Département d'Astrophysique, 91191 Gif-sur-Yvette, France; {\it bcordier@cea.fr}\\
%% Please give the E-mail address of the author, to whom future correspondence
%% requests will be sent.
        \and
            CEA/Paris-Saclay, Irfu/Département d'Électronique, des Détecteurs et d'Informatique pour la Physique, 91191, Gif-sur-Yvette, France;
        \and
             Centre National d’Etudes Spatiales, Centre Spatial de Toulouse,Toulouse Cedex 9, France;\\
\vs\no
   {\small Received 202x month day; accepted 202x month day}}

%% Authors should provide an abstract normally of not more than 200 words.
\abstract{
The scientific success of the SVOM mission will rely on the rapid transmission of alert messages from the satellite to the scientific community, and in particular to the ground-based instruments supporting the mission. In this paper, we present the alert system developed for SVOM, which relies on the rapid transmission of alert messages through the transfer of data packets from an onboard VHF-band radio transmitter to a network of radio receivers deployed along the satellite’s ground track.
We will successively detail the antenna design, radio performance, network deployment, its integration within the French data center, as well as the performance achieved after one year of operation in terms of availability and latency.
% Up to 6 keywords from this list: https://www.raa-journal.org/sub/author/keywords/index.html
}

   \authorrunning {B. Cordier, L. Jeannin, Ph. Lafabrie et al. }            %author_head in even pages
   \titlerunning{The SVOM VHF alert network}  % title_head in odd pages

   \maketitle
%% The author head (on even pages) and the title head (on odd pages) will be
%% automatically extracted from \author{} and \title{}. Whenever the title is too long,
%% you will be asked to supply a shorter one by inserting either \authorrunning{} or
%% \titlerunning{} before \maketitle. Anyway, you can specify your own heads.
%%
%%
%% Note: In the following text body of your manuscript, please note several differences from
%%       other major journals:
%% (1) \subsection{Please Capitalize the First Letter of Each Notional Word in Subsection Title}
%% (2) Please Capitalize the First Letter of Each Notional Word in all tables' captions

%
%________________________________________________ sections below
%
\section{Introduction}           %% first-level sections will be auto-capitalized
\label{sect:intro}

% and scientific objectives of the alert network

%% Authors can give a citation as 'Michel et al. 1992'.
%% You may also use \cite, \citep and \citet for citation, and use Table~1 or Figure~1
%% and so forth. Using \ref and \label for cross-references of Tables/Figures
%% is a good way in adjusting/adding/removing text, tables or figures.

SVOM is a French-Chinese mission dedicated to the detection and multi-wavelength study of the most distant stellar explosions: gamma-ray bursts (GRBs). The mission is composed of two sets of instruments: the first is integrated on board a satellite, while the second is located on Earth. A part of the satellite payload is designed to detect and localize the gamma-ray burst \citep{cordier+etal+2026}.

Since a GRB is a fleeting event, the scientific success of the mission relies on the ability to rapidly transmit localization data to the community, so that other ground-based or space-based instruments can provide follow-up observations and continue characterizing the transient event. One of the main objectives of this sequence, starting in space and continuing on the ground, is to measure the luminosity distance of the burst. This quantity can only be estimated with a large ground-based telescope equipped with spectroscopic capabilities in the infrared. To achieve this goal, it is crucial to alert these ground instruments as early as possible with the most accurate information available.

One of the SVOM mission requirements is to broadcast to ground based telescopes the measured GRB celestial coordinates in no more than 30 seconds after localization for 66 \% of the localized GRBs, and in no more than 20 minutes for 95\% of the GRBs. This requirement guided the entire design of the VHF alert network.

The SVOM alert network has been specifically designed to meet the requirement of rapid and continuous data transmission between the satellite and the ground. This transmission relies on the use of radio waves: the selected approach consists in integrating a radio transmitter on board the satellite and deploying, along its ground track, a set of radio receivers. These receivers, referred to as VHF stations, ensure the reception of the signal transmitted by the satellite, its decoding, and its conversion into digital packets. These packets are then forwarded via the internet to the French Science Center (FSC) in near real time \citep{Louvin+etal+2026}. Once received by the FSC, these digital packets are processed in order to extract the relevant information. This information is then used to generate and distribute notifications to the scientific community as well as to ground-based tracking instruments.

\begin{figure*}
\centering
\includegraphics[width=\textwidth, angle=0]{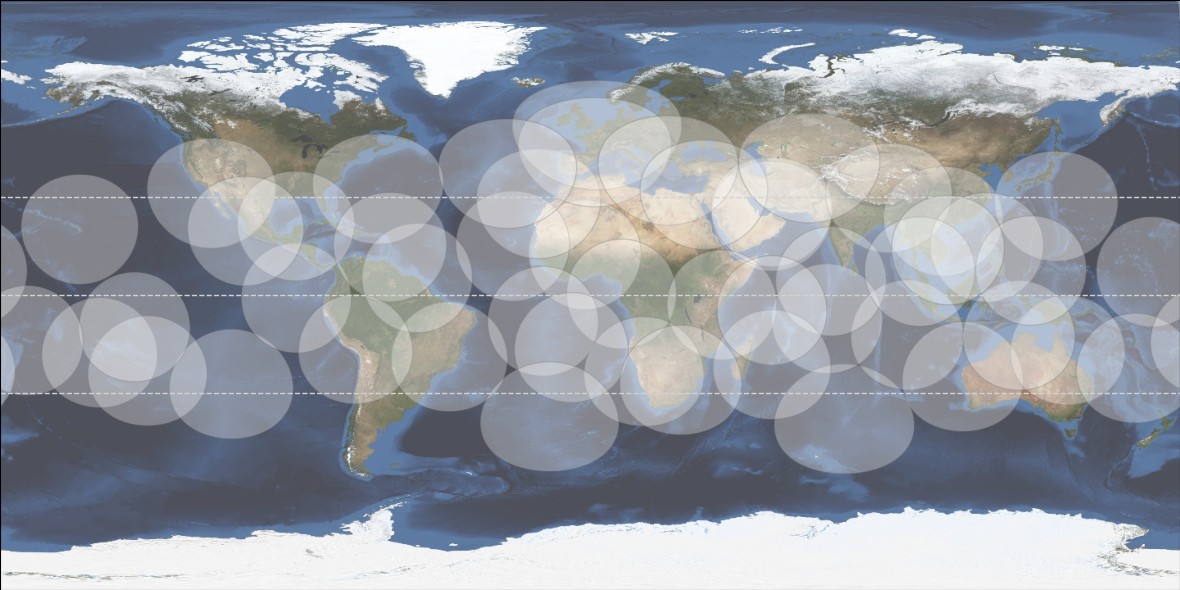}
   \caption{VHF network map (circles are drawn from a satellite elevation of 10° and a satellite altitude of 650 km)}
   \label{fig:Networkmap}
   \end{figure*}

SVOM is a low-Earth orbit satellite placed at an altitude of 650 km, completing one orbit around the Earth in just over 95 minutes, for about 15 revolutions per day. Its orbit is inclined at 30° with respect to the Earth’s equatorial plane. As a result, the satellite’s trajectory oscillates between latitudes –30° and +30°.
%Based on a minimum acquisition elevation of 10°, we estimate that the network should consist of at least 40 stations. To ensure sufficient redundancy and coverage in hard-to-reach areas, we have chosen to deploy a network of 50 stations, see Figure~\ref{fig:Networkmap}.
To receive the radio signal from the satellite, we have deployed around fifty radio receiver (exactly 53 as of November 7, 2025) along the satellite’s ground track, distributed as uniformly as possible across the Earth’s surface within this latitude range. These antennas make it possible to continuously receive the radio signal sent by the satellite, both during routine transmission and when a burst is detected, see Figure~\ref{fig:Networkmap}.

%Figure : Map of the network =================================
%%\begin{figure}[hptb]  

In the past, the HETE-2 space mission, which focused on Gamma-Ray Bursts, had also developed an alert system based on the transmission of radio messages and their reception on the ground by VHF stations \citep{Matsuoka+etal+2004}. The SVOM mission was largely inspired by this example but developed its own system. It is worth noting that, since the HETE-2 mission was in a nearly equatorial orbit, the number of stations to be deployed (around fifteen) was much smaller.
%Takahashi, T., et al. (2003). Prompt Gamma-Ray Burst Alert System of the HETE-2 Spacecraft. In: Proceedings of the 28th International Cosmic Ray Conference (ICRC 2003), Tsukuba, Japan, Vol. 5, p. 2741.

The following sections detail the station architecture, the expected radio performance, the deployment strategy for the different stations, the pre-launch testing procedures, and finally, the performance measured after one year of operation in orbit. In conclusion, we demonstrate that the SVOM VHF alert network fully satisfies the mission requirement.

\section{The VHF station}

\label{sect:station}
\begin{figure}[H]% ==> images de la station à remplacer  (texte et fleches ....)
    \centering
    \includegraphics[width= 0.4\textwidth]{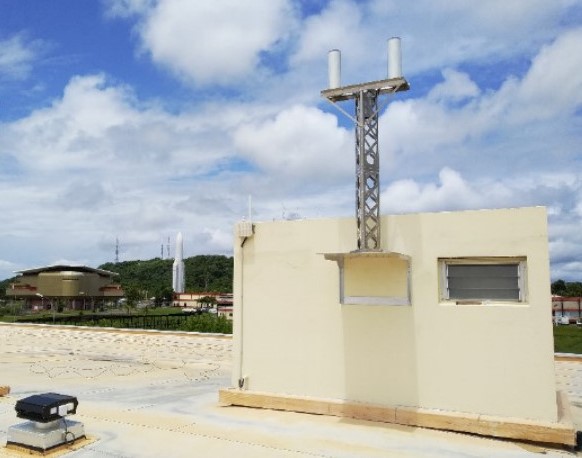}
    \caption{VHF operational ground station in Kourou}
    \label{fig:placeholder}
\end{figure}

A VHF station is composed of four main elements:
\begin{enumerate}
    \item two antennas to receive the SVOM emitted signal;
    \item a Software Design Radio (SDR) based on radio-frequency front-end and transceiver, which filters and digitizes the signal;
    \item a processing programmable module, which demodulates and decodes the signal to transform the received radio message in digital packets;
    \item an internet communication module, which ensures the real-time transmission of these digital packets to the French Science Center (FSC).
\end{enumerate}

\begin{figure}[H]% ==> images de la station à remplacer  (texte et fleches ....)
    \centering
    \includegraphics[width=0.3\textwidth]{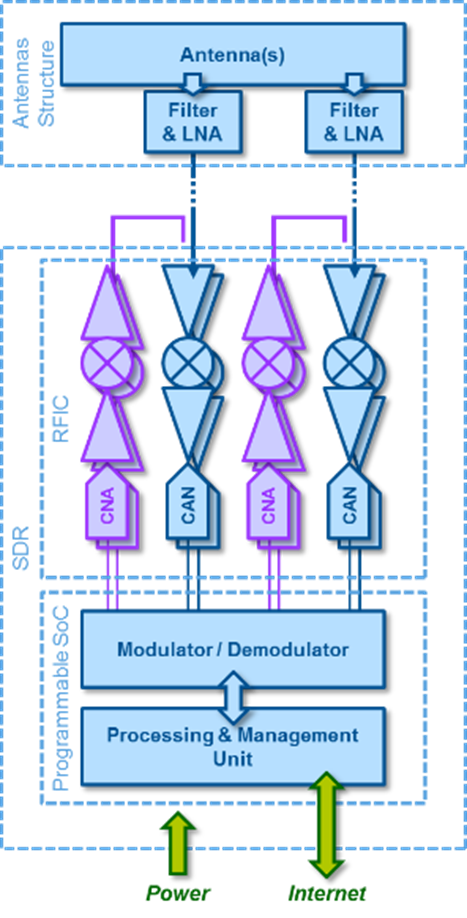}
    \caption{Ground station synoptic}
    \label{fig:placeholder}
\end{figure}

The key design drivers for the SVOM station were:
\begin{itemize}
    \item reasonable production costs;
    \item ease of assembly and installation;
    \item straightforward integration with the CNES secure network;
    \item reduced weight to minimize transportation costs.
\end{itemize}
The company INGESPACE, based in the Toulouse area, was selected by CNES to develop the SVOM station according to project specifications.

Given the number of stations to be deployed (50), a ``low-cost'' station concept was developed in order to minimize the overall cost of implementing the alert network. Otherwise, the number of stations required, coupled with the diversity of deployment sites and constraints would have made the total completion cost prohibitive.

Particular attention was given to the installation procedures, so that local site personnel could carry out the station installation themselves without the physical presence of a CNES technical agent, requiring only phone support if necessary. It should be noted that 80\% of the stations were deployed and installed during the Covid period.

With all of theses constraints the design chosen by INGESPACE consist on a very flexible and efficient SDR architecture with a full remote management and upgrade capacity. The SDR receiver is packaged on a small hardened sealed metal box with only  Radio Frequency (RF) connectors for antennas, and a command/control connector for internet connection to the FSC. 

On the mechanical side, a range of lightweight yet robust aluminum mounts was designed to adapt to the various configurations encountered at the deployment sites. Finally, the complete station is packaged in a transport case with a total weight of 40 kg.

\subsection{The Antennas}

The station is equipped with two quadrifilar antennas, operating in both Left-Hand Circular Polarization (LHCP) and Right-Hand Circular Polarization (RHCP). This configuration ensures the capacity of catching the SVOM signal for all geometrical configurations. %for the on ground parts (Azimuth / Elevation) and on board part (SVOM space craft attitude)

\subsection{Frequency and Modulation Scheme Choice}

The selection of the RF frequency band to ensure the reliability of the alert link was a complex task that needed to account for several factors, including:
\begin{itemize}
    \item the overall performance of the ground station system;
    \item applicable regulatory constraints; 
    \item the long-term sustainability of both performance and compliance throughout the mission lifetime.
\end{itemize}

Extensive studies have been conducted to assess the International Telecommunication Union (ITU) recommendations regarding the use of RF bands for space-to-ground communication in scientific missions. Several options have been identified, such as VHF, L-band, and S-band, each presenting specific technological, architectural, and environmental constraints.

\begin{comment}
The final chosen solution with the best trade of between 
\begin{itemize}
    \item RF band availability and sustainability
    \item Technological complexity for both on ground and on spacecraft parts 
    \item Costs to design deploy and maintain the system 

\end{itemize}

Is the space to Earth VHF band from 137 to 138 MHz witch have some global ITU protection in term of use, and a not too high frequency that allow not too complex electronic and antenna architecture with manageable low costs.
This Band is for example also used by some old NOAA spacecraft to broadcast real time weather pictures.
On this VHF band, contrary to old NOAA spacecraft, we have choose to implement a "modern" numeric modulation and coding solution to get a robust and spectral efficient signal transmission from the SVOM spacecraft to the ground network.

The chosen modulation scheme is a 4-CPFSK modulation (4 Continuous Phase Frequency Shift Keying) consisting on coding a 2 bits with a four state symbol discrete frequencies.
\end{comment}

The final solution was selected as the best compromise between:
\begin{itemize}
    \item the availability and long-term sustainability of the RF band;
    \item the technological complexity of both the ground and spacecraft segments;
    \item the overall costs of system design, deployment, and maintenance.
\end{itemize}

The chosen option is the space-to-Earth VHF band from 137 to 138~MHz, which benefits from global ITU protection for its use. This frequency range is sufficiently low to enable relatively simple electronic and antenna architectures, while keeping implementation costs manageable. For instance, this band has historically been used by NOAA satellites to transmit real-time meteorological images.

Unlike the legacy NOAA spacecraft, the SVOM mission adopts a modern digital modulation and coding scheme to achieve robust and spectrally efficient signal transmission from the spacecraft to the ground network. Specifically, the selected modulation is 4-CPFSK (4-level Continuous Phase Frequency Shift Keying), in which each symbol encodes two bits using four discrete frequency states \citep{Schonhoff+2003}.

\begin{figure}[h] 
    \centering
    \includegraphics[width=1\linewidth]{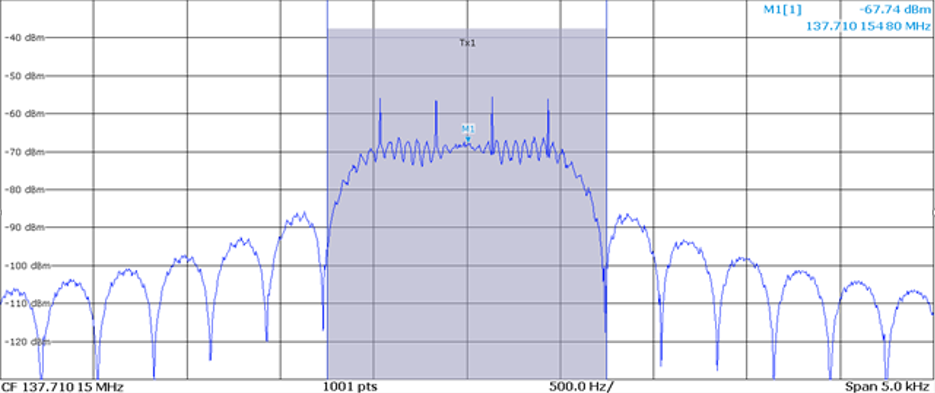}
    \caption{4-CPFSK Modulation scheme frequency spectrum (4 possible tones=2 bits by symbol)}
    \label{fig:placeholder}
\end{figure}

%A classic Reed-Solomon CCSDS channel coding has been added to the modulation to improve his robustness and introduce the capacity to add error correction capacity at ground level 
A classic Reed-Solomon code, based on CCSDS (Consultative Committee for Space Data Systems) standards, has been added to the modulation to enhance its robustness and enable error correction at ground level \citep{ReedSolomon+1960}. 
    
\subsection{Link Budget}

\begin{comment}
The alert system requirements don't need a high data rate to put the alert messages from the SVOM spacecraft to the ground network, a data rate of 600 bit/s has been fixed, giving a 300 symbol/s with the 4-CPFSK modulation chosen.
A link budget has been computed to get on the receiver an 6.9 dB of Eb/N0 (signal to noise ratio)  with an acceptable 7.10-3 bit error rate. To compute this link budget and produce a link specification, with ground and spacecraft requirements we have to manage some system constraints like spacecraft attitude, ground station position, but one of the most important consideration is the noise level on the ground  stations. as already stated the choose of 137-138 VHF ITU band is very accessible in term of technology and therefore easyly disturbed or jammed by many industrial or communication equipments.
\end{comment}

The alert system requirements do not demand a high data rate for transmitting alert messages from the SVOM spacecraft to the ground network. A data rate of 600~bit/s has been specified, corresponding to 300~symbols/s with the selected 4-CPFSK modulation.  

A link budget has been established to achieve an $E_b/N_0$ of 6.9~dB at the receiver, which ensures a bit error rate of approximately $7 \times 10^{-3}$, considered acceptable for this service. In deriving this link budget and producing a complete link specification, several system constraints must be taken into account, such as spacecraft attitude and ground station geometry. However, one of the most critical considerations is the noise environment at the ground stations. As previously discussed, the selected VHF ITU band (137--138~MHz) is technologically accessible but also highly susceptible to interference or jamming from various industrial and communication equipments.  

\subsubsection{Ground Jamming}

\begin{comment}
    Some studies and measurement campaign has been made to identify the source of disturbance and jamming on this theoretically reserved RF band. 
Two main families of ground disturbance has been identified: 
\begin{itemize}
    \item industrial noise 
    \subitem electronics devices like computers, power supplies, etc...
    \subitem communication devices like network devices with Ethernet links and cables 
    \item Ground communication noise with 
    \subitem ground talkie walkies 
    \subitem ground communication system for aircrafts and airports
    \subitem some "bad" saturated emitters on the VHF radio FM broadcasting on the 88.5-108 MHz 
\end{itemize}
A measurement campaign has been made in France with a mobile antenna to determine a best case far from cities and industrial zone where the noise is the lower, a middle case where the noise is acceptable and a worst case where the noise is maximum, inside cities or very near industrial facilities or communication buildings 

The results of the measurement campaign give the three hypothesis below for the link budget computations:

    \begin{itemize}
        \item 1 000 K as best case far from cities and industrial zone where the noise is the lower
        \item 3 000 K as middle case where the noise is acceptable
        \item 10 000 K as worst case where the noise is maximum, inside cities or very near industrial facilities or communication buildings
    \end{itemize}
\end{comment}

Several studies and measurement campaigns have been conducted to identify sources of disturbance and jamming within this theoretically reserved RF band. Two main categories of ground-based interference have been identified:
\begin{itemize}
    \item industrial noises, including:
    \begin{itemize}
        \item electronic devices such as computers, power supplies, etc.;
        \item communication equipment such as network devices with Ethernet links and cables;
    \end{itemize}
    \item ground communication noises, including:
    \begin{itemize}
        \item handheld radio;
        \item ground communication systems for aircraft and airport operations;
        \item improperly configured or saturated VHF FM broadcasting transmitters in the 88.5--108~MHz band.
    \end{itemize}
\end{itemize}

A dedicated measurement campaign was carried out in France using a mobile antenna to characterize the noise environment. Three representative scenarios were identified: a best-case scenario far from cities and industrial zones, where the noise level is minimal; an intermediate case, where the noise remains acceptable; and a worst-case scenario, where noise levels are highest, typically in urban areas or in close proximity to industrial facilities and communication infrastructures.

Based on this campaign, the following hypotheses were adopted for link budget computations:
\begin{itemize}
    \item 1\,000~K for the best-case scenario (remote areas with minimal noise);
    \item 3\,000~K for the intermediate case (moderate noise levels);
    \item 10\,000~K for the worst-case scenario (urban or industrial environments with maximum noise).
\end{itemize}

\subsubsection{Link disturbances}
\begin{comment}
On the path of the VHF signal between the SVOM spacecraft and the ground stations, the signal is likely to be subject to some atmospheric disturbance or absorptions. Thanks to his very long wavelength the VHF signal is very robust to tropospheric of H2O from clouds and rain, but this frequency is subject to RF absorptions form scintillation phenomena on ionosphere zone, The scintillation effect occurs on the low latitude zones mainly during the night, it can produce a massive absorption of 10 dB of the signal on local zone an for a limited time. Some studies on L Band around 1 GHz cans be used to get some hypothesis of occurrences for the link budget. 
\end{comment}
    
Along the propagation path of the VHF signal between the SVOM spacecraft and the ground stations, the signal may be affected by atmospheric disturbances and absorption phenomena. Owing to its long wavelength, the VHF signal is highly robust against tropospheric attenuation caused by water vapor in clouds and rain. However, this frequency range is susceptible to radio-frequency absorption due to ionospheric scintillation. Such scintillation effects occur predominantly at low latitudes, especially during nighttime, and can lead to localized signal fades of up to 10~dB for limited durations.

Studies conducted in the L-band region around 1~GHz provide useful references for estimating the occurrence probability of these scintillation events, and can therefore be used to support the definition of link budget assumptions \citep{AolBuchertJurua2020}.

\subsubsection{Ground station RF Interface}

\begin{comment}
At ground level for the ground stations, The RF interfaces requirements has been also set to get a at minimum elevation a minimum gain of -15 dBi for all azimuth directions with two antennas in opposite circular polarization.
\end{comment}

At ground level, the RF interface requirement for the stations is to have a gain of at least -15~dB at minimum elevation in all azimuth directions. This is achieved through the use of two antennas operating in opposite circular polarizations.

\subsubsection{Link budget operating point and synthesis}
\begin{comment}
    The modulation scheme, the coding of the channel and the noise hypothesis being posed, the space craft RF interfaces has been specified to get a good link for the worst case geometrical configuration for lower visibility elevation and worst case onboard VHF antenna gain. A midlle case for the noise has been choose to compute the spacecraft RF interfaces requirements. An EIRP between -8 and +4 dBW has been fixed for the spacecraft, with a 4PI STERADIAN radiation pattern for the two antennas in opposite circular polarization.
\end{comment}

With the modulation scheme, channel coding, and noise hypotheses defined, the spacecraft RF interfaces have been specified to ensure a reliable link under the worst-case geometrical configuration, namely low elevation angles and minimum onboard VHF antenna gain. For the computation of the spacecraft RF interface requirements, the intermediate noise scenario was selected. The spacecraft effective isotropic radiated power (EIRP) has been set between -8 and +4~dBW, assuming a $4\pi$~steradian radiation pattern achieved by two antennas operating in opposite circular polarizations.

\section{Station deployment}
\label{sect:deployment}

\subsection{Ensuring Global Coverage}

One of the major challenges of the project is to ensure the fastest possible alert transmission to the ground, regardless of the satellite’s position in its orbit. Even though the satellite operates only between -30° and +30° latitude, this region covers a vast portion of the Earth's surface, with particularly large oceanic areas where installing ground stations is extremely difficult, if not impossible.

%In this configuration, every second counts:
The alerts must be transmitted to the ground immediately upon detection, without waiting for the satellite to pass over a station. This necessitates a dense and longitudinally well-distributed ground network, extending even to areas that are often overlooked, such as remote islands, politically sensitive regions, and technically challenging locations. In particular, the world’s oceans—the Atlantic, Indian, and especially the Pacific—represent significant potential coverage gaps if not explicitly addressed.

\begin{comment}
To fill these gaps, several solutions are being considered or already implemented: deployment of stations on strategic islands, local partnerships, and the use of satellite connections (such as VSAT or Starlink). These measures maximize the likelihood that, wherever the satellite is, it can immediately transmit its alert to a reachable station, thus minimizing the delay in ground reception.
\end{comment}

\subsubsection{Maintaining satellite visibility}

Ensuring satellite visibility from the ground is a critical requirement for the immediate transmission of data, particularly  critical alerts. For a satellite in a low equatorial orbit such as SVOM, this visibility is inherently brief and intermittent, limited to a few minutes during each pass over a station. This necessitates not only extensive geographic coverage but also an optimal distribution of stations so that at least one is always able to ``see'' the satellite.

The challenge is therefore not merely to deploy stations, but to do so in the right locations. Each station covers only a limited visibility cone, depending on its latitude, the satellite’s altitude, and the local topography. Poor placement, insufficient elevation, or obstructed horizons can significantly reduce network performance. To address these constraints, precise orbital simulations were used to identify the most relevant areas and maximize the total cumulative visibility over a 24-hour period.

In parallel, the resilience of the network also relies on redundancy: if a station becomes unavailable (due to network outages, weather, or maintenance), another station must be able to take over without any loss of coverage. This ultimately ensures continuous satellite visibility and guarantees that every alert, regardless of its location along the orbit, can be transmitted immediately to the ground.

\subsubsection{Finding host sites}

Identifying suitable host sites for ground stations is one of the most complex phases of the project. Beyond purely technical constraints, this stage involves human, geopolitical, and operational considerations. The region between -30° and +30° latitude spans highly heterogeneous countries: some offer reliable infrastructure, a clear administrative framework, and access to connectivity, while others present major challenges in terms of stability, communication, or environment.

The priority is to identify local partners capable of hosting a station under safe, stable, and remotely accessible conditions. These partners may include universities, observatories, research institutes, or existing infrastructures already accustomed to working with space or scientific technologies. However, such collaborations take time, involving administrative negotiations, agreement signings, regulatory compliance, equipment import logistics, and local staff training.

In certain remote areas, as in Samoa or on Easter Island, an alternative solution is to deploy a fully autonomous station, powered by solar energy and connected to the internet via satellite. While this type of deployment increases technical complexity, it enables coverage of otherwise inaccessible regions. Regional or international partnerships may also facilitate access to strategic territories, particularly islands or regions with high orbital coverage value.

%Thus, finding host sites requires balancing technical opportunities with field realities, while always keeping in mind the ultimate goal: ensuring readiness to receive and immediately transmit an alert.

\subsubsection{Host site selection process}

The selection of a site to host a VHF station is not limited to its ideal geographic position. It depends on a multitude of technical, practical, and regulatory criteria, some of which are absolutely critical to ensure reliable and stable radio reception.

One major, often underestimated criterion is the level of radio frequency (RF) noise in the immediate environment. %Although the VHF band is effective for satellite-to-ground communication, it is particularly vulnerable to local interference. The presence of nearby RF emission sources—such as FM antennas, telecom relays, solar inverters, poorly shielded computer equipment, etc., can saturate the band or generate intermittent disturbances, making reception difficult or even impossible.
For this reason, every potential site must undergo an RF survey using a wideband receiver or spectrum analyzer coupled with a calibrated VHF antenna. These measurements allow the detection of existing interference and the identification of ``quiet'' areas, often far from dense urban zones, technology hubs, or industrial areas.

This factor is particularly critical because the satellite signals received in VHF are extremely weak, sometimes barely above the background noise. A clean electromagnetic environment is therefore essential to ensure stable reception, even under unfavorable conditions (low elevation, distant satellite).
This criterion adds to other key parameters in site selection:
\begin{itemize}
    \item clear horizon mask;
    \item ease of maintenance;
    \item reliable internet connection;
    \item logistical and administrative accessibility;
    \item clear regulatory framework, particularly regarding VHF reception authorization.
\end{itemize}

Taken together, these constraints require a precise and methodical selection process, sometimes at the expense of sites that are geographically attractive on paper. A fully functional station in a slightly less central location, requiring minimal maintenance support, is preferable to an ideal site on paper that is effectively blind due to excessive RF noise.

\subsection{SVOM Network deployment}

The deployment of the station network began in 2019, building on a solid foundation: the existing CNES network, which already included several strategically distributed ground stations across the globe. This initial backbone provided partial coverage and a proven architecture onto which the project could be rapidly integrated.

To extend the network, particularly within the equatorial zone, the project team leveraged an existing scientific collaboration network established over many years. Thanks to close ties with laboratories, universities, and observatories, new stations could be installed at reduced cost, often co-located with already existing infrastructure. These partnerships not only minimized hosting expenses but also provided access to local expertise for equipment monitoring and maintenance.

In more remote or difficult-to-access regions—such as isolated islands, underserved tropical zones, or politically sensitive contexts—deployment relied on technical or institutional partners already present on site. These partners played a key role by managing logistics, setting up the stations, performing first-level maintenance, and in some cases acting as intermediaries for VHF licensing or customs procedures. This collaborative model proved essential to accessing critical coverage areas without excessively increasing costs or deployment timelines.
The network currently consists of 50 stations. The list of the different stations with their characteristics is presented in the Appendix~\ref{app:Station lists}.
Today, the network continues to expand following this hybrid approach: capitalizing on existing resources, developing scientific partnerships, and relying on local intermediaries in regions with low accessibility. This agile and collaborative strategy has proven effective while also fostering international scientific cooperation around the project.

\subsubsection{Coverage}

The alert network currently provides coverage of 93\% of the orbit (93\% of the telemetry emitted by the satellite can be received at the FSC). In practice, the network’s coverage is even better than theoretical estimates, as SVOM receiving stations often acquire the telemetry signal at elevations below 5°.

%Figure : Coverage of the network =================================
%%\begin{figure}[hptb]  
\begin{figure*}
\centering
\includegraphics[width=\textwidth, angle=0]{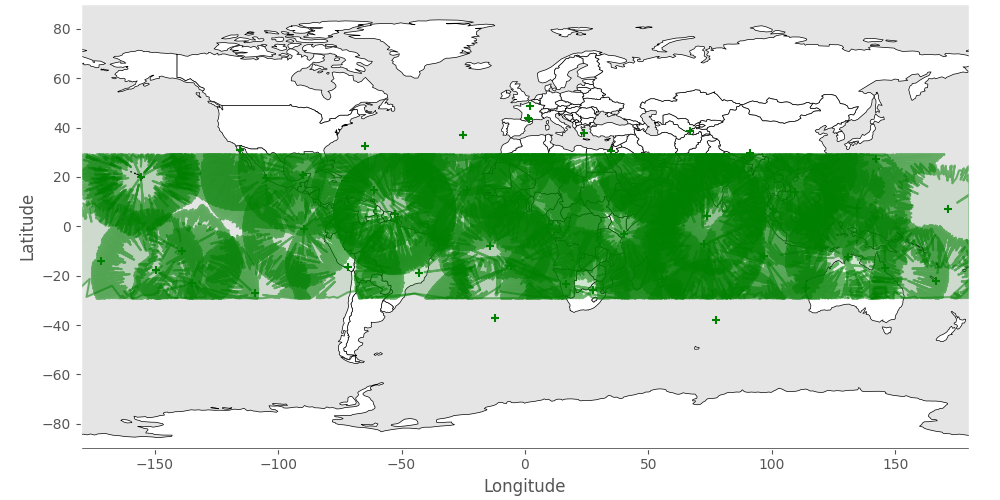}
   \caption{VHF network coverage constructed from packets received by each station. For a given station, at each pass, the position of the satellite is plotted for the first and last packets received.}
   \label{fig:coverage}
\end{figure*}   

% Table : coverage results
\begin {table*} 
\centering
\begin{tabular}{p{4cm} p{4cm} p{4cm}}
\hline
\centering\textbf{Number of station} & 
\centering\textbf{\% of the network covered by exactly N stations} & 
\centering\textbf{\% of the network covered by at least N stations} \tabularnewline
\hline
\centering 0 & \centering 7.3 & \centering - \tabularnewline
\centering 1 & \centering 18.8 & \centering 92.7\tabularnewline
\centering 2 & \centering 30.0 & \centering 73.7\tabularnewline
\centering 3 & \centering 23.3 & \centering 43.7\tabularnewline
\centering 4 & \centering 14.0 & \centering 20.4\tabularnewline
\centering 5 & \centering 5.3 & \centering 6.4\tabularnewline
\hline
\end{tabular}
\caption{Percentage Coverage Based on the Number of Stations Considered as of December 5, 2025. These Percentages are Measured from the Packets Received by Each Station at the Visibility Limit.}
\label{tab:tab_coverage}
\end{table*}

Figure~\ref{fig:coverage} shows the actual geometric coverage of the network, derived from the packets received by each station. The uncovered areas can be quickly identified and are mainly located in the North Pacific. Table~\ref{tab:tab_coverage} provides quantitative measurements of the coverage as a function of the number of stations considered. The uncovered area represents 7.3\% of the total coverage, and roughly three quarters of the area overflown by SVOM is within visibility of at least two VHF stations.

%\section{GROUND STATION NETWORK MAINTENANCE}
%\label{sect:maintenance}
%\subsection{SCHEDULED VS ACTUAL MAINTENANCE}

\subsection{Problems Encountered During the Deployment}
\label{sect:problems_encontered}

The deployment of the ground station network was not free from technical and operational challenges. Several major issues were encountered during the successive deployments, requiring significant adaptations to ensure the stability and performance of the system.

\subsubsection{Antenna-related issues}

Thanks to measurements taken with NOAA satellites (see section~\ref{sec:NoAA}), we were able to identify a mechanical design flaw in the initially deployed antennas. Excessive thermal expansion, particularly in tropical climates, caused progressive damage to the internal structure of the antennas, leading to gain loss and malfunctions. In 2023, all antennas in the network were replaced with a new generation of antennas that are more resistant to temperature variations, ensuring stable long-term performance.

\subsubsection{Network connectivity issues}

Several sites experienced difficulties communicating through the internet, with connections that were highly unstable or extremely limited in bandwidth. In some cases, governmental restrictions made establishing a VPN to CNES nearly impossible. With the support of the CNES network team, specific solutions were implemented: routing via local relays, protocol optimization, and adaptation of systems to operate under very low data rates, while maintaining the ability to transmit critical alerts.

\subsubsection{Electromagnetic (EMC) issues}

The sensitivity of the VHF system to electromagnetic noise proved particularly problematic at certain sites. The metallic structure of the station itself sometimes act as a resonant element, re-emitting disturbances onto the antennas, which then picked up parasitic signals present on the horizon. To address this issue, several measures were implemented:
\begin{itemize}
    \item additional shielding of coaxial cables;
    \item improved cable routing within the structure to avoid ground loops;
    \item installation of a ground plane beneath the antenna to reduce coupling with the structure and reflections from the ground.
\end{itemize}
These modifications reduced the noise floor by approximately 10 dB, significantly improving reception quality in complex RF environments.

\subsection{Preliminary Tests with NOAA Satellites}
\label{sec:NoAA}

To test the network during the deployment phase (see Sect. \ref{sect:deployment}), the stations were configured to receive data from the polar-orbiting weather satellites NOAA-18 and NOAA-19, operated by the U.S. National Oceanic and Atmospheric Administration (NOAA) ~\citep{Goodrum2000}. With an orbital period of approximately 102 minutes, these satellites complete about 14 passes over the equator each day—both ascending and descending. As a result, the same region is typically overflown at least four times per day, at intervals of roughly six hours. Their respective frequencies are 137.9125 MHz and 137.10 MHz. They transmit analogue signals using the Automatic Picture Transmission (APT) system. Earth images are derived from this signal with a fixed width and a transmission rate of 120 lines per minute. The number of lines depends on the duration of the passage. The first line corresponds to the moment the satellite becomes visible from the station, and the last line to the moment it disappears. 

\begin{comment}
    Thanks to the ephemeris data from NOAA-18 and NOAA-19, and using the coordinates of a specific site, we were able to determine the start and end times of each satellite pass, as well as the start and end azimuths and the maximum elevation. Using this information, we matched the images to their corresponding passes. Knowing the time at which each image was written to disk, we calculated the elevation at the end of the pass. Then, using the transmission rate, we computed back the acquisition start time and the elevation at the beginning of the pass.
\end{comment}

Using the ephemeris data from NOAA-18 and NOAA-19, together with the coordinates of a given site, we determined the start and end times of each satellite pass, as well as the corresponding start and end azimuths and the maximum elevation. Based on this information, the images were matched to their respective passes. By referencing the timestamps associated with each image file, we estimated the elevation at the end of the pass. Finally, applying the transmission rate, we inferred the acquisition start time and the elevation at the beginning of the pass.

In the presence of obstructions (buildings, mountains) or, if the station sensitivity is too low, the station's horizon may differ from the ideal 0° elevation horizon. If the signal reception is poor, signal loss may occur, resulting in missing or noisy lines in the image, including missing lines at the beginning or end of the satellite pass.

Since the beginning of 2020 to the beginning of 2024, we monitored the evolution of the image quality to test how the stations evolves with the time. We also measured the real horizon of each station in order to derive the evolution of coverage throughout the deployment phase.

\subsubsection{Image quality assessment}

\begin{figure*}[ht!]
\begin{center}
\includegraphics[scale=0.7]{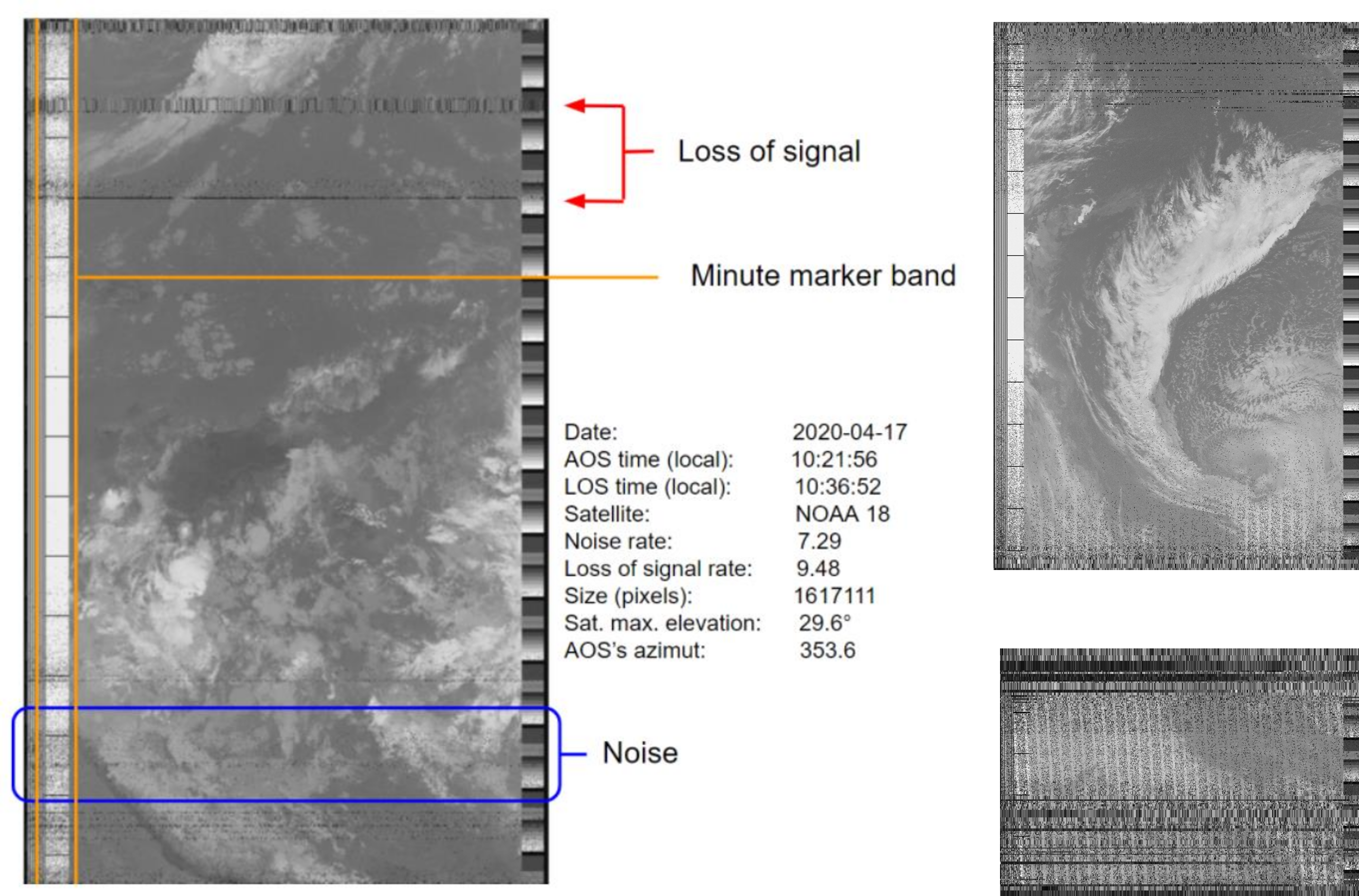}
\caption{On the left: the three parameters extracted from the image to construct the quality score: the number of lines of signal loss, the number of black pixels in the minute mark band, and the size of the image.
On the right, two images collected by the Azores station on January 7, 2024. Top: example of a good-quality image with a score of 37.5. Bottom: example of a poor-quality image with a score of 99.3.}
\label{fig:NOAA_1.png}
\end{center}
\end{figure*}

The quality of the images is assessed thanks to three criteria:
\begin{itemize}
    \item the size of the image which is directly linked with the horizon of the station;
    \item the number of repeated lines when there is a loss of signal (shown in red in Fig.~\ref{fig:NOAA_1.png});
    \item the noise that is measured by counting the ratio of black pixels within the minute marker band (shown in orange in Fig.~\ref{fig:NOAA_1.png}).
\end{itemize}

By linearly combining these three criteria, a global quality rating between 0 (best) and 100 (worst) is given to each image. As an example, Figure~\ref{fig:NOAA_1.png} provides two examples of images received by the Azores station (Portugal). The best image, rated 37.5,  has much more lines and fewer noisy ones than the worst image rated 99.3 (for the first week of year 2024).

% Est-ce qu'on pourrait combiner ces deux images en une seule?
\begin{comment}
  \begin{figure}[h!]
  \begin{minipage}[t]{0.45\textwidth}
  \centering
   \includegraphics[width=0.9\textwidth]{figures/imageSat_infra_20240106_124351.png}
	  \caption{Worst image from the Azores station (Portugal), received on 2024/01/06 (infrared NOAA channel)}
    \label{fig:worst_acores}
  \end{minipage}%
  \begin{minipage}[t]{0.45\textwidth}
  \centering
   \includegraphics[width=0.9\textwidth]{figures/imageSat_infra_20240107_001526.png}
	  \caption{Best image from the Azores station (Portugal), received on 2024/01/07 (infrared NOAA channel)}
    \label{fig:best_acores}
  \end{minipage}%
\end{figure}
\end{comment}

\subsubsection{Contribution of preliminary NOAAS tests}

With the images received from the stations, we ran recurrent analysis to monitor the quality of the reception (through the images quality ratings) over time.

\begin{figure}[!h]
\begin{center}
\includegraphics[width=0.55\textwidth]{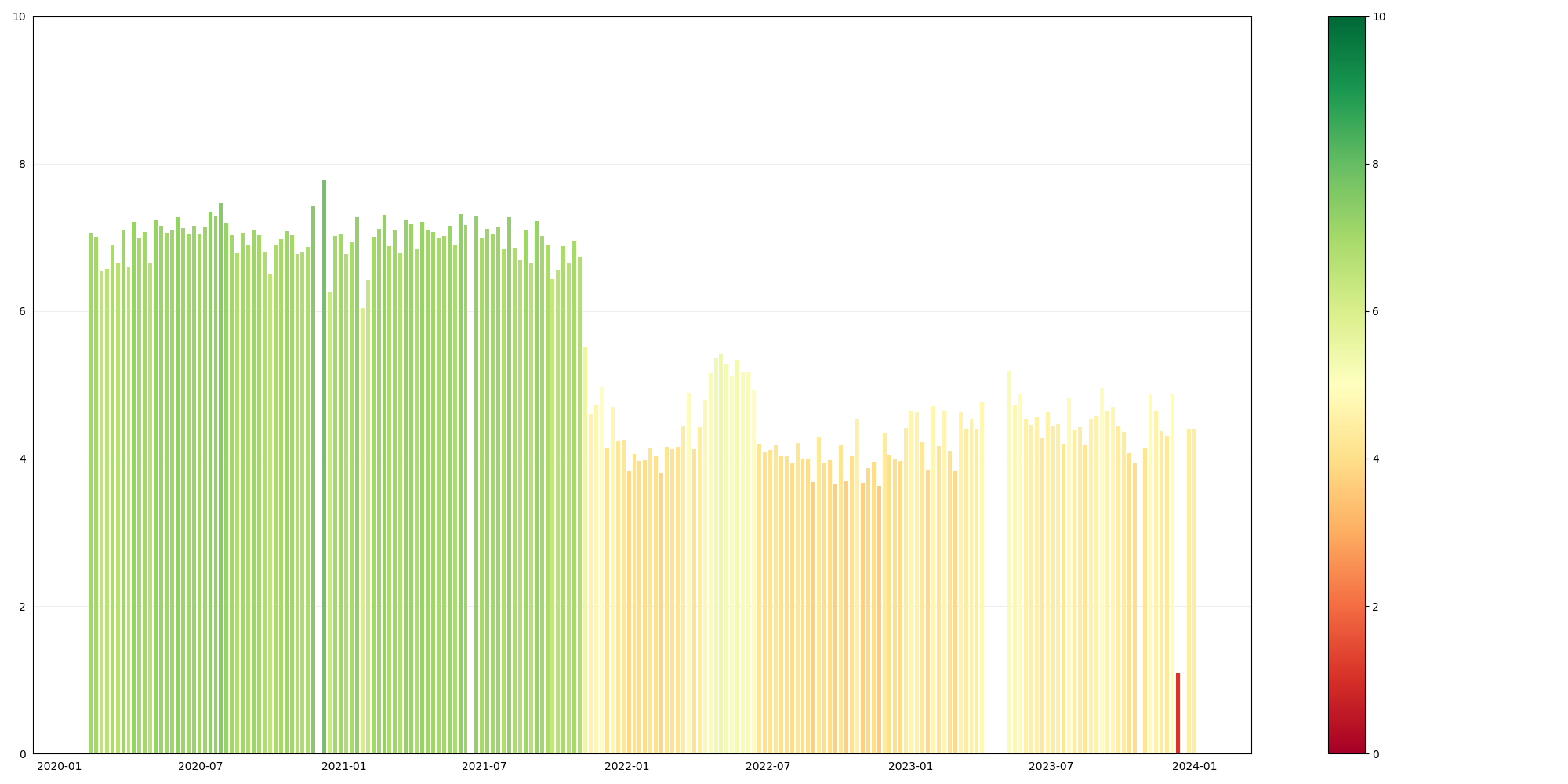}
\caption{Evolution of the mean quality rating of the NOAA images received by the Hartebeesthoek station (South Africa)}
\label{fig:notes_HBK}
\end{center}
\end{figure}

Figure~\ref{fig:notes_HBK} shows this evolution for the Hartebeesthoek station (South Africa). The evolution of the mean image rating illustrates effects that alter the quality of the reception (see Sect. \ref{sect:problems_encontered}). Gaps correspond to network issues. The fast degradation in Autumn 2021 is due to the premature ageing of the antenna. Despite a replacement by a new upgraded station in Spring 2022, the quality did not fully recover due to electromagnetic issues in the environment.

\section{FSC integration}
\label{sect:fsc}

VHF data received from ground antennas are sent to the French Science Center (FSC), which hosts all services for SVOM data processing. The FSC comprises approximately 40 software components, primarily running as micro services in a Docker Swarm cloud environment built on virtual machines managed via OpenStack \citep{Louvin+etal+2026}. The production infrastructure is hosted by CC-IN2P3 in Lyon (France). SVOM FSC services depend on external systems for:
\begin{itemize}
    \item data storage and classification (e.g., relational databases);
    \item monitoring (e.g., Grafana, Loki);
    \item authentication and authorization (Keycloak).
\end{itemize}

The relational database infrastructure is maintained by CC-IN2P3 administrators, while the other components are deployed in a separate cloud and are under the responsibility of CEA FSC administrators.

The VHF binary messages (packets) are received at FSC through a dedicated REST API by the so-called \textit{vhf-manager} service, which saves them in their raw format in a dedicated database. The \textit{vhf-manager} allows insertions of multiple packets within one HTTP request from a given ground station. 

The raw data format and rates of VHF packets are the described bellow.
\begin{itemize}
    \item Total packet size is 128 bytes, with 28 bytes added by the ground station itself and 100 bytes coming from the satellite.
    \item The 100 bytes are composed by a common section (packet header) of 6 bytes, and 94 bytes of data which will depend on the instrument. The packet header contains an identifier (APID) which allows to recognize the specific format of the 94 bytes. SVOM instruments have a total of 58 APIDs.  
    \item The emission frequency of VHF packets from the satellite is about 1 packet every 2 seconds.
    \item The reception rates depend on the number of stations which are in the visibility range of the satellite, and we can measure over the day rates going from 0.2 packets per second to more than 1 packet per second. A comprehensive analysis of the VHF data reception is reported in chapter 6 of this paper.
\end{itemize}

Even if the information from the satellite is always organized in individual packets, some APIDs have the information spread over a large number of packets. We can mention as an example the APIDs which are used to organize the light-curve content measured by ECLAIRs and GRM instruments. The information of a light-curve is organized in 2 APIDs (because of a different time resolution) composed by 42 and 22 packets each. 

The packet processing workflow is summarized in the following paragraphs. 

\subsection{Raw packets ingestion}

The raw packet is recorded, the station header and packet header are parsed and their content is organized in dedicated relational tables. The remaining 94 bytes are stored with an hash identifier which allows to detect duplicate packets (the same packet can be delivered by 2 different ground stations, or the satellite can repeat also relevant packets to increase the efficiency of their reception). 
The \textit{vhf-manager} then calls a dedicated \textit{decoder} service which is responsible for reading the content of the binary VHF packets and converting them in JSON format (this depends on the APID). The JSON VHF packets are also saved in the VHF database. 
Once the packet ingestion is over, the system starts notifying clients about the packet reception, by sending the information (in JSON format) to a centralized messaging system hosted at FSC. Clients may subscribe to the VHF data messaging queue and act depending on the packet content \citep{Louvin+etal+2026}. The FSC provides via the \textit{vhf-manager} an authenticated REST API allowing applications and authorized SVOM collaborators to access the VHF data (both raw and interpreted) stored in the VHF database.

For the first year of data taking, the total volume of the VHF database is about 100 GB in the Postgres database, for less than 10 GB of pure raw data. 

\subsection{Alert management}

% Some specific APIDs contain information about satellite triggers respect to the observation of a potential gamma ray burst. We call those packets ``alerts'', and they can be send by GRM and ECLAIRs instruments. The alert information is organized in an individual packet and they are frequently repeated by the satellite, thus minimizing the possibility to loose that information at FSC level. The alert packets will trigger an additional service deployed at FSC and devoted to the creation of a flag (the burst ID) which is then used by all scientific pipelines to quickly identify the VHF packets which were involved in the downstream products. 

Some specific APIDs contain information about satellite triggers related to the observation of a potential gamma ray burst by either the GRM or ECLAIRs instruments (or both). We call those packets ``alerts''. For each alert, the localization of the GRB candidate is contained in an single 94-bytes packet we call ``alert packet''. Each alert packet is sent several times by the satellite, to maximize the probability of receiving this data at FSC level. The alert packets trigger an additional service deployed at FSC and devoted to the creation of a flag (the burst ID). After the initial alert packet, the instruments on-board SVOM broadcast more observation data in a sequence of VHF packets dubbed the ``alert sequence``, which are all flagged using the burst ID for easy identification and processing by the FSC scientific pipelines. 

\subsection{Satellite position and attitude}

Other APIDs contain information about satellite position and attitude, which are relevant for downstream physics analysis. This information is extracted from those packets and organized in specific tables.

\subsection{From raw packets to science products}

%The notification system allows other applications to react to the VHF packets arrival. Several pipelines are devoted to more complex analysis of the data, possibly taking into account multiple packets. 
The FSC notification system allows other applications to react to the VHF packets arrival. Several processing pipelines are devoted to real-time VHF data analysis, combining data gathered from multiple packets. 

\section{In-flight performances measurements for the VHF alert network} 
\label{sect:tests}

%Noaa18 et 19, network coverage, stations horizon, quality
%First analysis= oct 2019
%Last analysis= dec 2012

\subsection{Tests with the SVOM satellite }
% Partie à réorganiser en mettant d'abord les tests avec SVOM puis EP
% et en regroupant les ilusstrations SVOM et EP dans la même figure (gain de place et comparaison plus facile). Actions Philippe :-)

An in-flight qualification measurement campaign has been run to verify the on board requirements of the SVOM spacecraft for the VHF link.

%Many tests have been done  with a on ground calibrated antenna on the KOUROU CNES facilities in French Guyana with the target  verify the VHF EIRP and the VHF antenna gain of the SVOM Spacecraft from many spectrum measurements.
Extensive testing was conducted using a ground-calibrated antenna at the CNES facilities in Kourou, French Guiana, with the objective of verifying the VHF EIRP and antenna gain of the SVOM spacecraft from a series of spectrum measurements.

%From some passes with a consistent geometrical diversity, hundred of spectrum has been measured and compute with a reverse link budget to obtain the EIRP on the spacecraft plane.
Across several passes with sufficient geometrical diversity, hundreds of spectra were recorded and processed using a reverse link-budget approach to retrieve the EIRP on the spacecraft-plane.
\begin{comment}
\begin{figure}[H]
    \centering
    \includegraphics[width=0.5\linewidth]{CNES VHF Calibrated Antenna.png}
    \caption{CNES French VHF calibrated antenna}
    \label{fig:placeholder}
\end{figure}

\begin{figure}[H]
    \centering
    \includegraphics[width=1\linewidth]{VHF spectrum to antenna gain.png}
    \caption{From spectrum measurements to antenna diagram: a reverse link budget computation}
    \label{fig:placeholder}
\end{figure}

After some hundred  spectrum measurements a reverse link budget computations has been done to verify the EIRP requirement.
For EP spacecraft, the measurements shows a lack of 20 dB of EIRP between requirements and measurements. these results has been confirmed by the operational results, where the VHF alert packets are mainly loss by the ground network because of the missing of correct EIRP on the  onboard spacecrafts  
\end{comment}

\begin{figure}[H]
    \centering
    \includegraphics[width=1.0\linewidth]{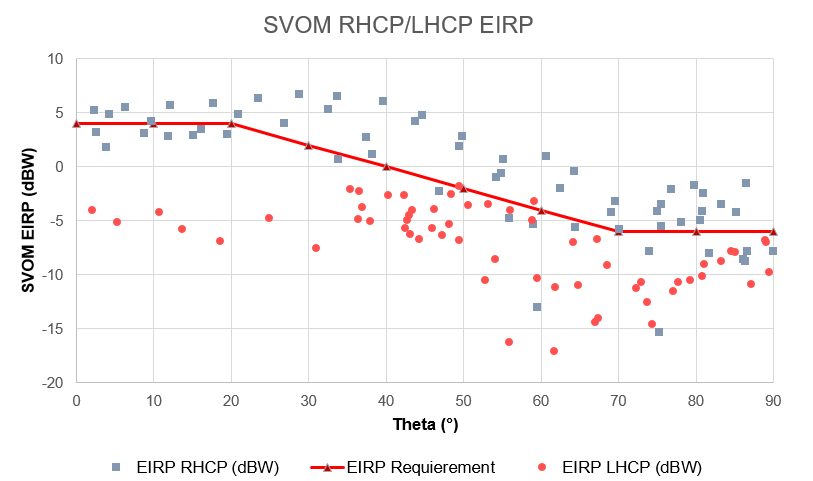}
    \caption{EIRP measurements and requirement in function of theta angle of the antennas (0° is on axis)}
    \label{fig:EIRP}
\end{figure}

The results presented in Figure~\ref{fig:EIRP} show that SVOM spacecraft reaches the EIRP requirement with his RHCP VHF antenna with a good margin but not with his LHCP antenna. The lower performances of the LHCP part is explained by a different location of the LHCP antenna on the spacecraft. Some scientific instruments and communication equipment near the LHCP antenna results on a disturbed radiated pattern on the LHCP antenna. 

\begin{figure}[H]
    \centering
    \includegraphics[width=0.85\linewidth]{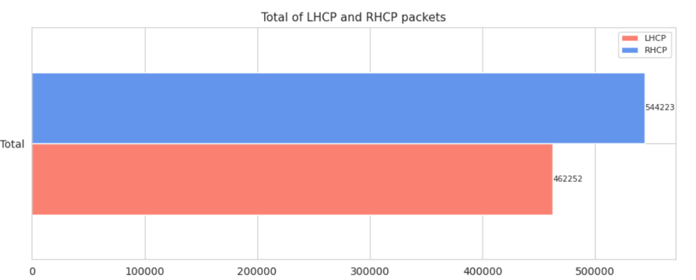}
    \caption{SVOM RHCP and LHCP packet received at FSC for a pass over the Kourou station.}
    \label{fig:RLHCP}
\end{figure}

% The better performances of the RHCP antenna on the satellite is clearly visible at the French science Center level, where more VHF packets are received from the RHCP antenna, see figure Figure~\ref{fig:RLHCP}.
The difference in performances between the antennas is clearly visible at the French science Center level, where more VHF packets are received from the RHCP antenna, see Figure~\ref{fig:RLHCP}.

From the EIRP measurement, a computation is made to obtain from the quaternion attitude of the spacecrafts the radiated pattern of both RHCP and LHCP antennas.

\begin{figure}[H]
    \centering
    \includegraphics[width=1.0\linewidth]{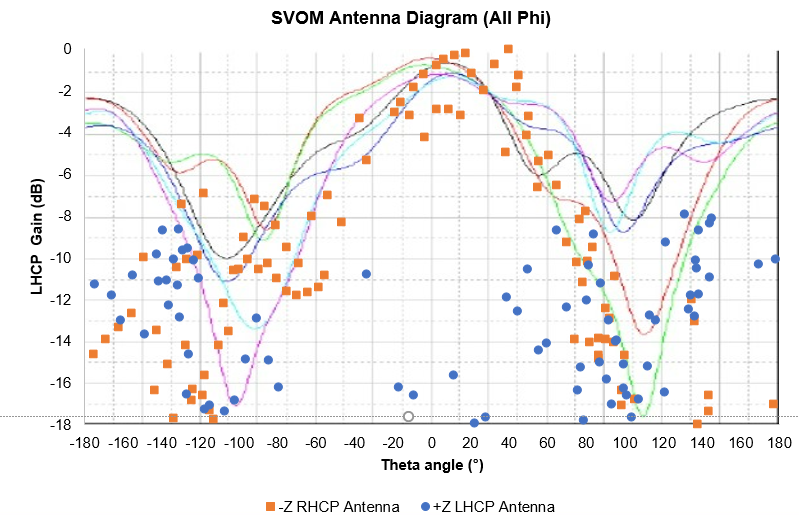}
    \caption{SVOM antenna gain in function of antenna theta angle (0° is on axis), continuous lines corresponds to before launch measurements, dots corresponds to in-flight measured values.}
    \label{fig:gain}
\end{figure}

%Computations for SVOM patterns shows the expected radiated pattern for the RHCP one and an on axis disturbed radiated pattern on the LHCP one.
Computations of the SVOM patterns show the expected radiated pattern for the RHCP antenna, whereas the LHCP antenna exhibits a perturbed on-axis radiation pattern, see figuree\ref{fig:gain}. We believe that this difference in results between the antennas is due to a different adaptation of RHCP and LHCP on the spacecraft.

%The tests show the feasibility to compute EIRP and gain diagrams of in-flight spacecraft from on ground spectrums RF measurement with a good calibrated antenna.

%The key points of this feasibility are a good accuracy and fine mesh on the calibration of the on ground antenna, typically 1°step in azimuth and elevation mesh. 

%A good accuracy on the time synchronization between the onboard quaternion and on ground RF measurement is also a major thing to ensure a good and accurate reverse link budget computation.

\begin{comment}
    Computations of EP VHF antenna gains show unexpected antenna gains that explain the end to end performances. The understanding of theses unexpected performances is under study by SECM.
\end{comment}

%\ Remonter la partie synthèse mesure au dessus et faire en 5.2 un focus rex SVOM

%\subsection{Tests conclusions}

%Many feedbacks can be made from theses in-flight tests of the VHF onboard tests.
%A major feedback of these tests is that for low frequency (VHF) links, the wavelength is relatively high and very sensitive to the placement of the antenna on the spacecraft, it is particularly visible on the %SVOM results with a dissymmetry on the results due to a different accommodation of the RHCP and LHCP on the spacecraft.

%Another main feedback on theses measurements is the strictly necessity to perform RF end to end tests in radiated mode, including the antennas to check the performances and avoid a lack of performances, even if the radiated tests are tricky to implement especially with these high wavelengths. 

\section{SVOM VHF end-to-end performance}
\label{sect:perf-end}

The SVOM VHF emitter has been turned on the day following SVOM launch in June 2024 and used continuously ever since as the main channel for real-time mission monitoring and alert broadcast. In this section we will present some results regarding the end-to-end performances of the VHF network during SVOM operations.

\subsection{Metrics and Methodology}
\label{sect:metrics}

For each VHF transmission, we define the following timestamps related to ground-to-board events:
\begin{itemize}
    \item the \textit{packet time}: date and time of the VHF TM creation on board by the instruments, encoded in the packet;
    \item the \textit{station time}: date and time at which a given station has received the VHF packet, added to the station header before transmission to FSC;
    \item the \textit{FSC reception time}: date and time at which the packet is saved at FSC, computed upon insertion in the dedicated VHF database.
\end{itemize}

Using those timestamps we compute two metrics: the ground transmission delay between the VHF antennas and the FSC which highlights the performances of the ground network, and the overall delay between the packets creations on board and their reception at FSC which is the actual measure of end-to-end performance. 

In addition to time-related metrics we also estimate the \textit{data completion} by computing the proportion of on-board VHF transmissions that have actually been received at FSC, using the list of all VHF packets sent by the on-board VHF emitter which is downloaded through X-Band communications every few hours alongside the complete SVOM observation data. 

\subsection{Repetition and duplication strategies}
\label{sect:repetition-duplication}

In order to ensure timely distribution of GRB detections and localization to the scientific community, all VHF messages containing alert-related data are sent several time by the satellite. This allows to circumvent holes in the VHF network coverage and punctual networking issues that could lead to packet loss. This is referred to as ``on-board repetition strategy''. The number and delay between repetitions depends on the packet types as identified by their APIDs.

To further improve the reliability of the SVOM downlink for alert-related data, some messages sent through the VHF network are also downloaded using the Chinese navigation satellite system BeiDou. This communication subsystem has a slower rate than VHF and a much smaller bandwidth, but as we will see in the next section it allows to speed up the on-ground reception of alert data in some cases, and could even be vitally important in case of VHF network temporary unavailability.

All VHF packets emitted by the SVOM that are not received in real-time through the VHF network or BeiDou are recovered every few hours alongside the complete SVOM observation data through the X-Band network which allows us to ensure completeness of VHF data with a few hours delay.

\subsection{Ground network performances}

The performance results presented in this section were computed using the data produced by SVOM during September 2025. This allows to analyze the performances of the system using the latest versions of hardware and software in place at the time of redaction.  

\begin{figure}[h]
    \centering
    \includegraphics[width=0.85\linewidth]{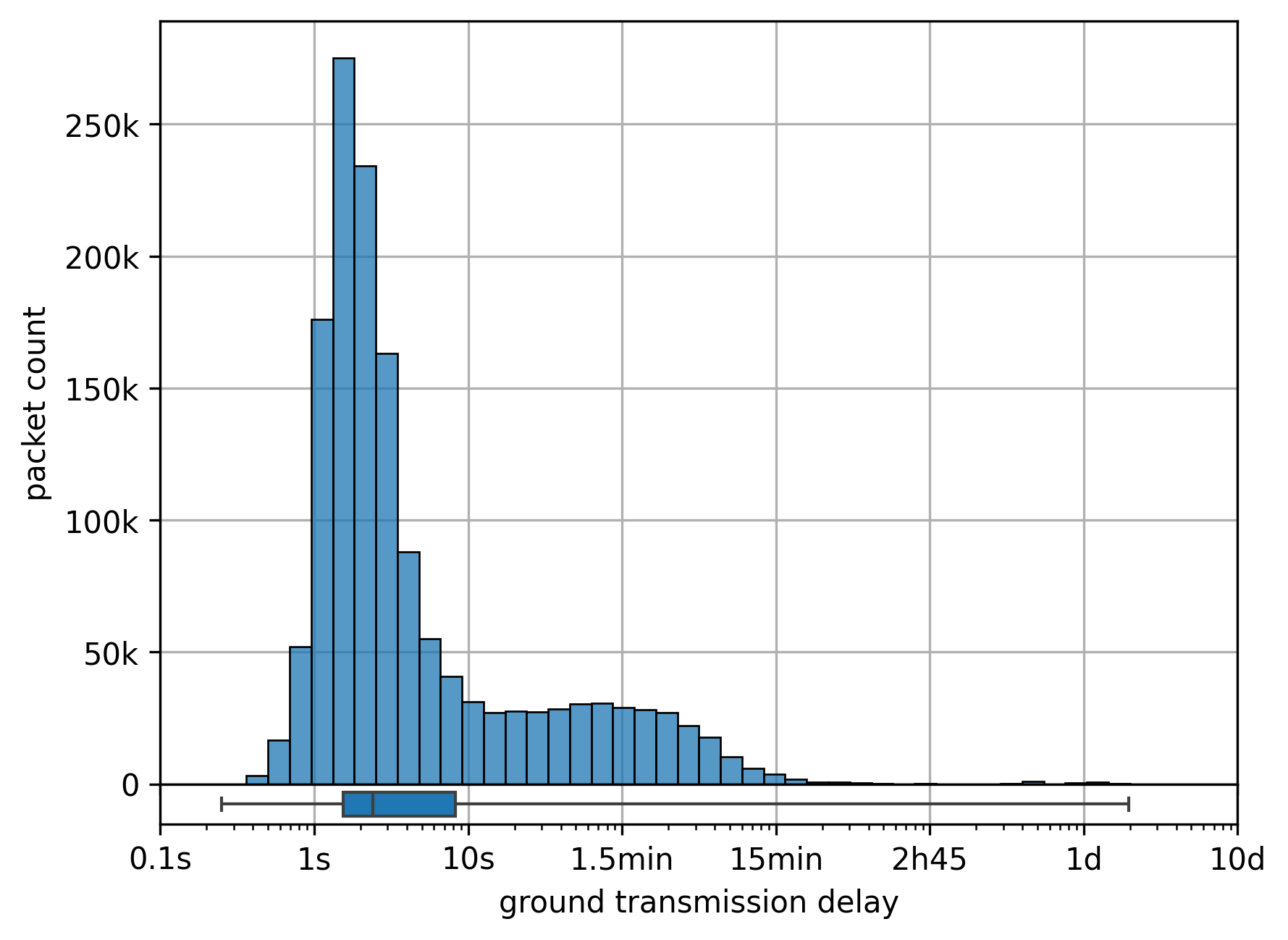}
    \caption{Distribution of stations-to-FSC transmission delays of SVOM packets received in Sep. 2025 by all stations}
    \label{fig:ground_delay_all}
\end{figure}

Figure~\ref{fig:ground_delay_all} shows" the distribution of ``ground transmission delays'' between VHF antennas reception and FSC saving time for the packets received in September 2025. The box plot at the bottom of the figure shows a median delay around 2.4 seconds. The box itself delimits the interquartile range representing 50\% of the data, meaning 25\% of the packets were transmitted by the stations in less than 1.5 seconds, and 75\% in less than 7 seconds.

\begin{figure}[h] 
    \centering
    \includegraphics[width=0.85\linewidth]{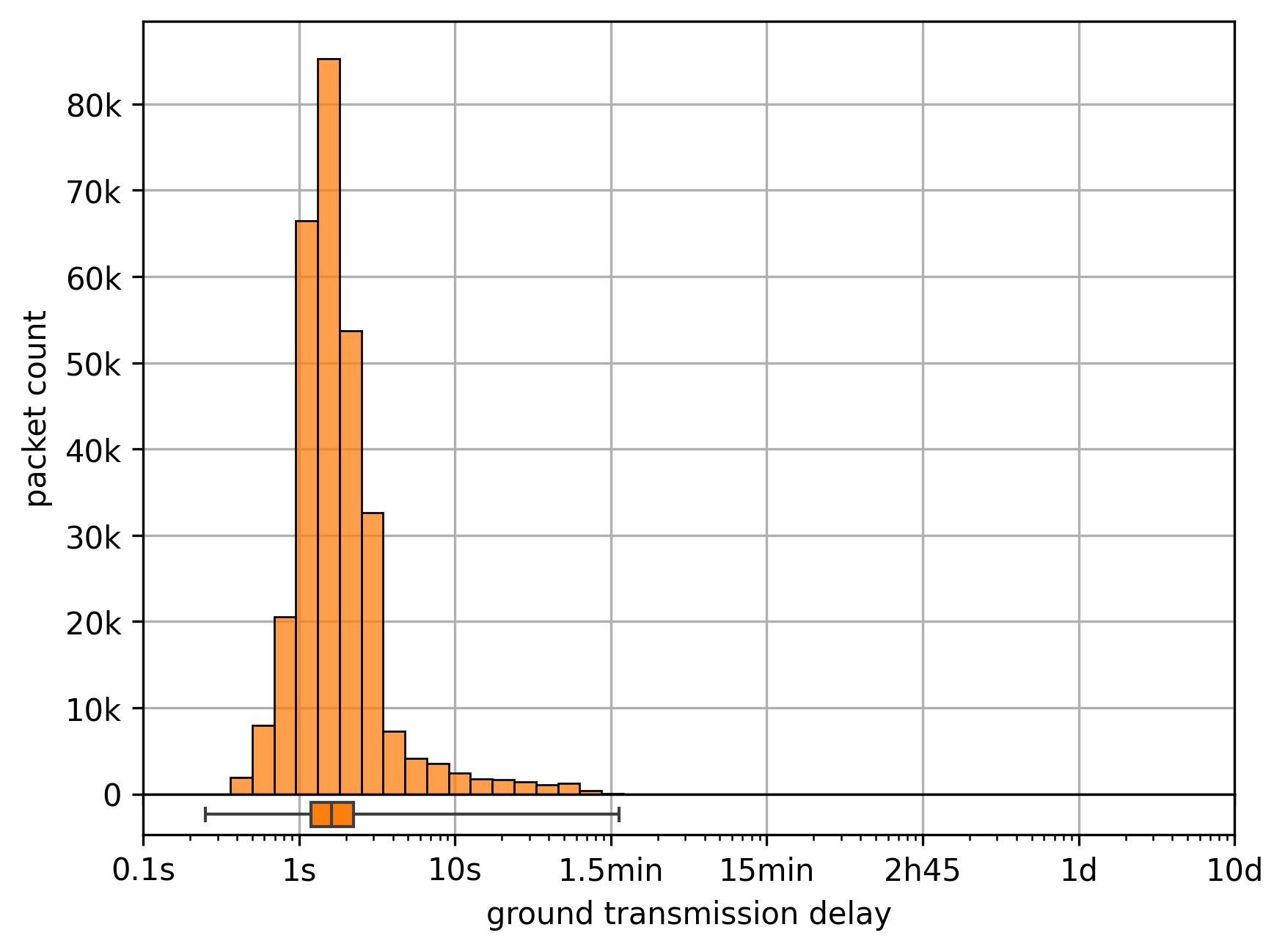}
    \caption{Distribution of stations-to-FSC transmission delays of SVOM packets received in Sep. 2025 by the 10 most efficient stations}
    \label{fig:ground_delay_best}
\end{figure}

In the vast majority of cases the performance of the ground network is very good. However we can see on Figure~\ref{fig:ground_delay_all} that there are sometimes significant communication delays between the VHF stations and FSC. Those delays are due to network and bandwidth issues in some VHF stations locations. In order to illustrate this, we show on Figure~\ref{fig:ground_delay_best} the same plot but restricted to the 10 most efficient VHF stations. We can see that the performances are excellent and the spread hugely reduced, with a median below 2 seconds, more than 75\% of the packets transmitted in less than 3 seconds and an overall maximum delay around 1 minute and 30 seconds.

Updates in the VHF stations software as well as optimizations in the communications between the VHF stations and FSC have been developed and are currently being deployed. Those are expected to significantly reduce the ground communication delays especially for stations which suffer from low bandwidth and internet speed. Those improvements should be implemented on all stations in the first half of 2026.

\subsection{End-to-end performances}

The performance results presented in this section were computed using the data produced by SVOM during September 2025. This allows to analyze the performances of the system using the latest versions of hardware and software in place at the time of redaction.

In Appendix B, the table presents the contribution of the different stations to the SVOM alert system. The table is ordered in decreasing order according to the number of packets received at the FSC from each station. An inhomogeneity in performance can be observed, which can be explained by a combination of several factors: the geographical location of the station, its sensitivity as determined by the local VHF environment, and the bandwidth of its internet connection.

Because the satellite is often visible from multiple stations, a given packet can be received at the FSC from several stations.The second column indicates, for each station, the number of its packets that arrived first.
 
Finally, the third column indicates the number of packets for which the station was the sole transmitter. This column shows that even if a station does not contribute significantly to the overall packet volume, it may fill a coverage gap in the network when it is the only station in communication with the satellite.
This table clearly illustrates the complementarity of the stations within the network, balancing performance and coverage.

Regarding latencies and completion of the overall system, the figures presented below show on the left-hand side a box-and-whiskers representation of the board-to-ground delays distribution for all SVOM communication channels: VHF, BeiDou and X-Band (see section \ref{sect:repetition-duplication}). Those delays correspond to the difference between the on-board message production and its reception and saving at FSC. The right-hand side bar plot shows the data completion for each of the communication channels for the data corresponding to the left-hand side distributions.

\begin{figure*}[ht] 
    \centering
    \includegraphics[width=\linewidth]{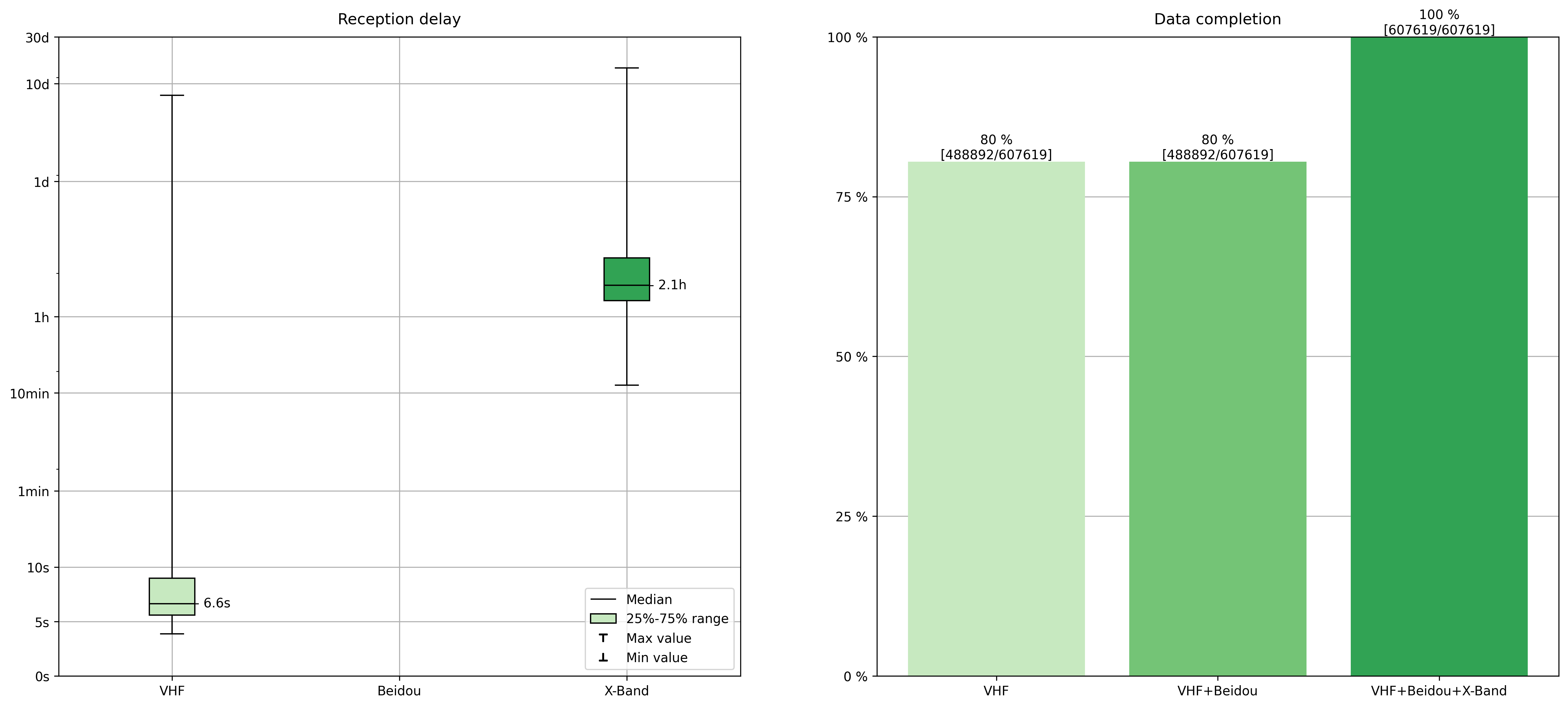}
    \caption{SVOM-to-FSC transmission delays and data completion for recurrent packets received in Sep. 2025}
    \label{fig:board_delay_rec}
\end{figure*}

\begin{figure*}[h] 
    \centering
    \includegraphics[width=\linewidth]{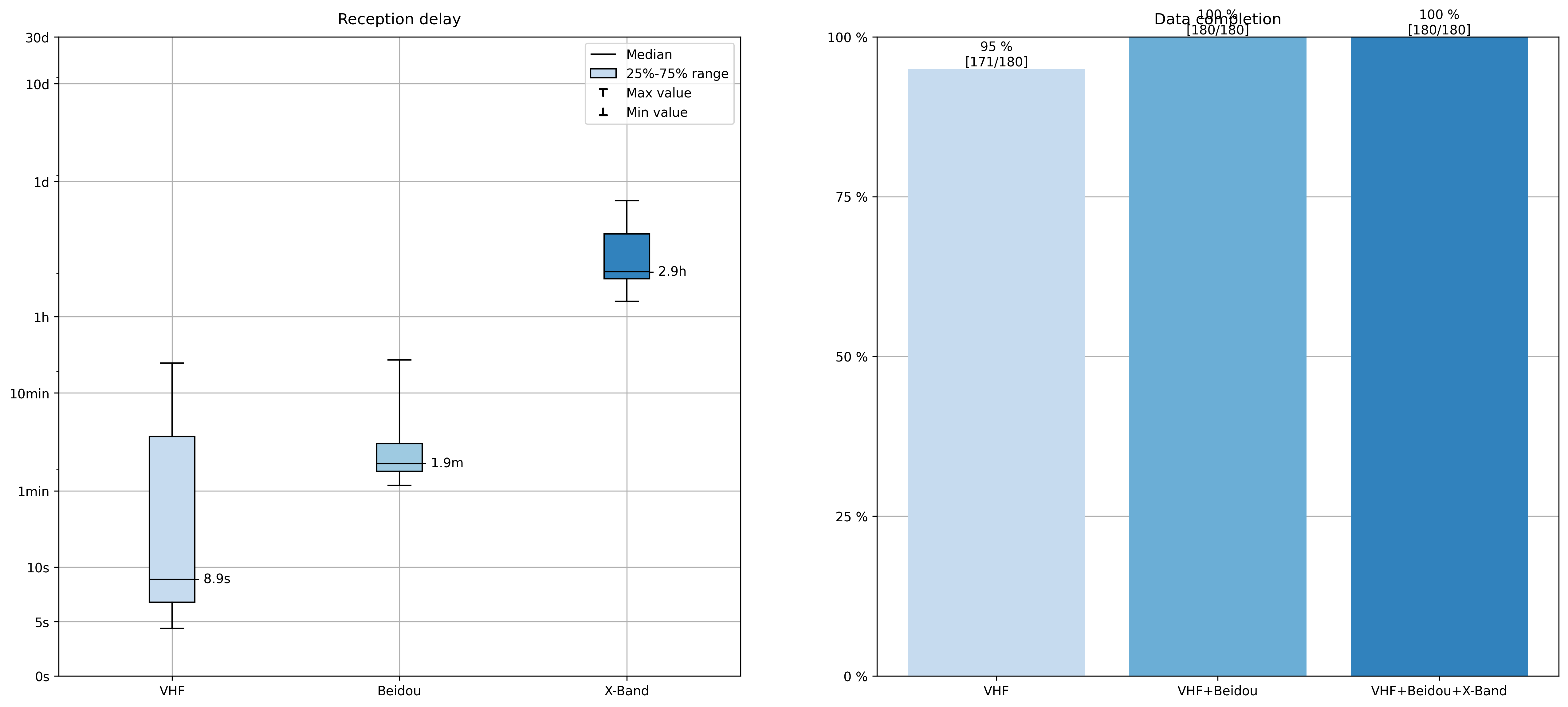}
    \caption{SVOM-to-FSC transmission delays and data completion for alert packets received in Sep. 2025}
    \label{fig:board_delay_alerts}
\end{figure*}

Figure~\ref{fig:board_delay_rec} shows the board-to-ground delays distribution and data completion over September 2025 restricted to the type of VHF packets that are downlinked by SVOM as soon as they are produced and that are never repeated by the on-board emitter nor duplicated through BeiDou. Those messages are mostly used to broadcast SVOM instruments monitoring messages. Studying the delays and completion for those packets allows to see the raw performance of the SVOM VHF communications without repetition strategy and BeiDou replication. We can see that the delay distribution is consistent with Figure~\ref{fig:ground_delay_all} with a shift of around 4 seconds due to the on-board processing and VHF spacecraft-to-station communication time. The data completion is of about 80\% meaning 20\% of packets are not received in the initial VHF transmission and only recovered through X-Band which median time is around 2 hours. These performances are sufficient for real-time monitoring but not quite good enough to comply with the scientific goals of the SVOM mission.

Figure~\ref{fig:board_delay_alerts} shows the board-to-ground delays distribution and data completion over September 2025 restricted to VHF packets containing on-board trigger data. Those messages are repeated several time by the on-board emitter and also replicated through the BeiDou network, and are used at FSC to automatically generate scientific products and notices which are broadcasted to the worldwide scientific community in real time. On the 180 on-board triggers that occurred in September 2025, we can see that:
\begin{enumerate}
    \item for the VHF network, the following alert delay distributions are measured: 25\% in less than 5.3 s, 50\% in less than 8.9 s, and 75\% in less than 215 s;
    \item the on-board repetition strategy causes a spread of the time delay distribution which makes sense since the delay between packet production and reception increases with each repetition of a VHF message;
    \item the on-board repetition strategy allows to raise the data completion through VHF alone to 95\%;
    \item the BeiDou replication has a median distribution delay of around 2 minutes which means that in some cases the Beidou replicated message arrives at FSC before the alert coming from the VHF communication;
    \item the BeiDou replication allows to raise the data completion to 100\% meaning we did not lose any VHF alert packet in September and never had to wait for the X-Band complete data to arrive to produce and broadcast SVOM alerts.
\end{enumerate}
The latency performance of the SVOM VHF network is comparable, even slightly better for more than half of the messages, to that of the TDRSS system used by the Swift satellite. According to information given by the swift team, TDRSS provides a median latency of 20\,s (14\,s / 37\,s for the 25\% / 75\%) in the last few years for broadcasting the first alert message, which includes the high-energy localization of the burst. In the case of the GECAM satellite, which uses the Beidou system, latencies of about 2 minutes are observed for the first alert message.

However, it should be noted that these messaging systems, which rely on satellite relay networks, can transmit only a limited number of messages and therefore do not allow continuous contact with the scientific mission. In contrast, the SVOM VHF network enables the continuous transmission of recurrent messages that track the status of both the instruments and the spacecraft platform. This capability is a major advantage for a mission focused on the transient sky, as it allows verification of instrument operating conditions at the time of the trigger and during follow-up observations. Such information is essential for the scientific validation of the content of the alert.

Finally, taking into account the repetition and duplication strategy, 63\% (113/180) of the first alert packets notifying on-board triggers were delivered in less than 30 seconds after the event, and 100\% in less than 7 minutes. We can therefore conclude that the SVOM alert system complies with the mission requirement outlined in the introduction to this article.

\section{Conclusions}
\label{sect:conclusion}

The development of the VHF alert network and the first In-Flight results of SVOM mission clearly show:
\begin{itemize}
    \item the system requirement about the alert distribution delays are fully satisfied;
    \item performances at station level shows that the ageing of main parts of ground station (antennas and receiver) is better than expected after one year of operation;
    \item SVOM alert network should allow to perform SVOM nominal and extended operational phases.
\end{itemize}
After one year of operation, and considering both the network's performance and the scientific results obtained, it is clearly demonstrated that the VHF alert system plays a central and irreplaceable role in the daily functioning and scientific operations of the SVOM mission.

\begin{acknowledgements}

The Space-based multi-band astronomical Variable Objects Monitor (SVOM) is a joint Chinese-French mission led by the Chinese National Space Administration (CNSA), the French Space Agency (CNES), and the Chinese Academy of Sciences (CAS). We gratefully acknowledge the unwavering support of NSSC, IAMCAS, XIOPM, NAOC, IHEP, CNES, CEA, and CNRS.

We would like to thank the trainee engineers from École Centrale Lille/Lyon who worked at CEA on monitoring the VHF network: Sarah Menouni, Hugo Thiesse, Aubry De Longcamp, Elie Straumann, and Lucas Witzel Ferreira.
\end{acknowledgements}

\bibliography{bibtex}

\appendix
\onecolumn
\section{VHF stations List}
\label{app:Station lists}
\small
\begin{longtable}{|c|l|l|l|l|l|l|l|}
\hline
\textbf{N°} & \textbf{ID} & \textbf{Name} & \textbf{Country} & \textbf{Installation date} & \textbf{Latitude} & \textbf{Longitude} & \textbf{Implementation}  \\
\hline
\endfirsthead

\hline
\textbf{N°} & \textbf{ID} & \textbf{Name} & \textbf{Country} & \textbf{Installation date} & \textbf{Latitude} & \textbf{Longitude} & \textbf{Implementation} \\
\hline
\endhead

1  & ZAF1  & Hartebeesthoek & South Africa & 2019-04-29 & -25.89 & 27.71 & CNES network \\
2  & GUF1  & Kourou & France & 2019-05-20 & 5.17 & -52.69 & CNES network \\
3  & GUF2  & Kourou 2 & France & 2019-05-20 & 5.17 & -52.69 & CNES network \\
4  & GRC1  & Athens & Greece & 2019-07-15 & 37.98 & 23.78 & Scientific collaboration \\
5  & GAB1  & Libreville & Gabon & 2019-09-12 & 0.39 & 9.60 & CNES network \\
6  & SHN2  & Saint Helena & UK & 2019-11-19 & -15.94 & -5.67 & Agreement \\
7  & THA2  & Songkhla & Thailand & 2020-01-15 & 7.16 & 100.61 & Scientific collaboration \\
8  & REU1  & La Réunion & France & 2020-01-22 & -21.20 & 55.57 & Scientific collaboration \\
9  & PHL1  & Manila & Philippines & 2020-01-23 & 14.54 & 121.04 & Agreement \\
10 & BFA1  & Ouagadougou & Burkina Faso & 2020-04-02 & 12.37 & -1.51 & Agreement \\
11 & KEN1  & Malindi & Kenya & 2020-04-02 & -3.00 & 40.19 & CNES network \\
12 & CPV2  & Praia & Cape Verde & 2020-08-31 & 14.92 & -23.51 & Agreement \\
13 & ISR1  & Tel Aviv & Israel & 2020-09-10 & 30.60 & 34.76 & Scientific collaboration \\
14 & SHN1  & Ascension & UK & 2020-09-25 & -7.92 & -14.33 & CNES network \\
15 & ESP2  & Maspalomas & Spain & 2020-10-06 & 27.76 & -15.63 & Agreement \\
16 & MTQ1  & Le Lamentin & France & 2020-10-31 & 14.60 & -61.00 & Agreement \\
17 & ARE3  & Sharjah & UAE & 2020-11-05 & 25.31 & 55.49 & Scientific collaboration \\
18 & PYF2  & Papeete & France & 2021-02-08 & -17.57 & -149.60 & Scientific collaboration \\
19 & PYF3  & Rikitea & France & 2021-03-15 & -23.13 & -134.96 & Agreement \\
20 & AUS3  & Carnarvon & Australia & 2021-06-11 & -24.87 & 113.70 & Agreement \\
21 & PRT1  & Açores & Portugal & 2021-06-30 & 36.99 & -25.13 & Agreement \\
22 & DGA1  & Diego Garcia & UK & 2021-07-05 & -7.27 & 72.38 & Agreement \\
23 & ATF1  & Amsterdam & France & 2021-09-13 & -37.80 & 77.57 & Agreement \\
24 & FR4   & CNES Fermat & France & 2021-09-13 & 43.56 & 1.48 & CNES network \\
25 & SHN3  & Tristan Da Cunha & UK & 2021-10-27 & -37.07 & -12.31 & Agreement \\
26 & ARE1  & Al Aïn & UAE & 2021-11-25 & 24.12 & 55.41 & Scientific collaboration \\
27 & CHN1  & Nanning & China & 2021-11-26 & 22.51 & 107.78 & Scientific collaboration \\
28 & SYC1  & Mahe & Seychelles & 2021-12-06 & -4.68 & 55.53 & Agreement \\
29 & USA2  & Hawaii & USA & 2021-12-13 & 20.02 & -155.67 & Scientific collaboration \\
30 & UZB1  & Maidanak & Uzbekistan & 2022-02-26 & 38.67 & 66.90 & Scientific collaboration \\
31 & ECU2  & Galapagos & Ecuador & 2022-04-19 & -0.90 & -89.61 & Agreement \\
32 & PLW1  & Palau & Palau & 2022-04-29 & 7.46 & 134.53 & Agreement \\
33 & MEX1  & San Pedro Martir & Mexico & 2022-06-07 & 31.05 & -115.47 & Scientific collaboration \\
34 & BMU1  & Bermuda & Bermuda & 2022-06-30 & 32.35 & -64.66 & CNES network \\
35 & NCL1  & Nouméa & France & 2022-07-25 & -22.28 & 166.46 & Agreement \\
36 & JPN1  & Chichi-jima & Japan & 2022-10-12 & 27.09 & 142.19 & Scientific collaboration \\
37 & PER1  & Arequipa & Peru & 2022-10-15 & -16.47 & -71.49 & Agreement \\
38 & PYF4  & Marquises & France & 2022-11-14 & -9.78 & -139.01 & Agreement \\
39 & FR3   & CEA & France & 2023-03-10 & 48.73 & 2.15 & Scientific collaboration \\
40 & MEX3  & Chamela & Mexico & 2023-04-11 & 19.50 & -105.04 & Scientific collaboration \\
41 & AUS11 & Coco Islands & Australia & 2023-04-20 & -12.19 & 96.83 & Agreement \\
42 & MEX2  & Yucatán & Mexico & 2023-05-11 & 20.99 & -89.73 & Scientific collaboration \\
43 & CHN3  & Lhassa & Chine & 2023-05-31 & 29.64 & 91.18 & Scientific collaboration \\
44 & AUS7  & Darwin & Australia & 2023-07-06 & -12.37 & 130.87 & Agreement \\
45 & AUS6  & Cairns & Australia & 2023-08-02 & -16.82 & 145.68 & Agreement \\
46 & CHL8  & Easter Island & Chile & 2023-11-02 & -27.14 & -109.41 & Agreement \\
47 & MDV1  & Male & Maldives & 2024-01-25 & 4.19 & 73.53 & Scientific collaboration \\
48 & BRA8  & Guanhaes & Brasil & 2024-08-13 & -18.79 & -42.92 & Agreement \\
49 & NAM1  & Windhoek & Namibia & 2024-10-22 & -23.27 & 16.50 & Scientific collaboration \\
50 & VNM3  & Quy Nhon & Vietnam & 2025-01-13 & 13.72 & 109.21 & Scientific collaboration \\
51 & USA3  & Hawaii & USA & 2025-04-02 & 20.02 & -155.67 & Scientific collaboration \\
52 & WSM1  & Apia & Samoa & 2025-10-25 & -13.85 & -171.79 & Agreement \\
53 & MHL1  & Majuro & Marshall Islands& 2025-11-07 & 7.09 & 171.38 & Agreement \\
\hline
\end{longtable}

\clearpage
\section{Number of VHF packets received per station at the French Science Center from September 1 to 30, 2025.}
\label{app:Performance}
In September 2025, the stations in Sharjah, Amsterdam, CNES Fermat, and Chichi-jima were not operational, and the stations in Apia and Majuro had not yet been installed. The stations are listed in descending order of packets received at the FSC.

% lignes fines
\setlength{\arrayrulewidth}{0.3pt}

% espacement vertical compact
\renewcommand{\arraystretch}{1.1}

% colonnes largeur fixe centrées verticalement + gauche
\newcolumntype{L}[1]{>{\raggedright\arraybackslash}m{#1}}

% colonnes largeur fixe centrées verticalement + horizontalement
\newcolumntype{C}[1]{>{\centering\arraybackslash}m{#1}}
\small
\setlength{\tabcolsep}{4pt}

\begin{longtable}{
|L{0.7cm}
|L{1.3cm}
|L{3.0cm}
|C{2.6cm}
|C{2.6cm}
|C{2.6cm}|
}
\hline
\textbf{N°} &
\textbf{ID} &
\textbf{Name} &
\textbf{SVOM packets received from this station} &
\textbf{SVOM packets received first from this station} &
\textbf{SVOM packets received only from this station} \\
\hline
\endfirsthead

\hline
\textbf{N°} &
\textbf{ID} &
\textbf{Name} &
\textbf{SVOM packets received from this station} &
\textbf{SVOM packets received first from this station} &
\textbf{SVOM packets received only from this station} \\
\hline
\endhead

1 & NCL1 & Nouméa & 46387 & 40502 & 26429 \\
2 & MTQ1 & Le Lamentin & 40633 & 9587 & 8378 \\
3 & AUS11 & Coco Island & 40083 & 28353 & 19828 \\
4 & PYF2 & Papette & 39019 & 30929 & 17880 \\
5 & REU1 & La Réunion & 38849 & 18607 & 8221 \\
6 & MEX3 & Chamela & 35863 & 22331 & 14459 \\
7 & ESP2 & Maspalomas & 34747 & 23835 & 2763 \\
8 & ZAF1 & Hartebeesthoek & 34315 & 3091 & 1298 \\
9 & SHN2 & Saint Helena & 33806 & 20911 & 2498 \\
10 & PYF3 & Rikitea & 32319 & 6074 & 5565 \\
11 & CPV2 & Praia & 32071 & 13655 & 1657 \\
12 & DGA1 & Diego Garcia & 30740 & 11233 & 1965 \\
13 & NAM1 & Windhoek & 30649 & 13477 & 719 \\
14 & THA2 & Songkhla & 29420 & 20298 & 2581 \\
15 & PYF4 & Marquises & 27471 & 17474 & 4134 \\
16 & ECU2 & Galapagos & 27040 & 17996 & 14586 \\
17 & ISR1 & Tel Aviv & 26886 & 20093 & 8365 \\
18 & AUS7 & Darwin & 26637 & 15304 & 13167 \\
19 & AUS6 & Cairn & 26320 & 11337 & 3348 \\
20 & MDV1 & Male & 25414 & 15015 & 4575 \\
21 & USA2 & Hawai & 25246 & 5611 & 4324 \\
22 & CHN1 & Nanning & 24753 & 7727 & 4848 \\
23 & SHN1 & Ascension & 24563 & 2246 & 2225 \\
24 & USA3 & HawaI & 24521 & 22994 & 3866 \\
25 & GUF1 & Kourou & 24405 & 8053 & 956 \\
26 & GAB1 & Libreville & 21930 & 14047 & 3246 \\
27 & BRA8 & Gunahaes & 21264 & 17543 & 13964 \\
28 & GUF2 & Kourou 2 & 21261 & 12495 & 521 \\
29 & PHL1 & Manila & 21002 & 13656 & 4187 \\
30 & MEX1 & San Pedro Martir & 20722 & 16539 & 5885 \\
31 & BFA1 & Ouagadougou & 18565 & 3819 & 941 \\
32 & PRT1 & Açores & 17119 & 5488 & 485 \\
33 & SHN3 & Tristan Da Cunha & 16719 & 4435 & 1840 \\
34 & CHL8 & Easter Island & 16176 & 13270 & 6257 \\
35 & PLW1 & Palau & 15479 & 8077 & 6155 \\
36 & SYC1 & Mahe & 13943 & 9910 & 1376 \\
37 & KEN1 & Malindi & 13929 & 8731 & 3128 \\
38 & BMU1 & Bermuda & 12886 & 12116 & 1387 \\
39 & CHN3 & Lhassa & 12402 & 2401 & 2299 \\
40 & GRC1 & Athen & 12350 & 8601 & 630 \\
41 & VNM3 & Quy Nhon & 11873 & 4082 & 296 \\
42 & ARE1 & Al Aïn & 10874 & 4325 & 2672 \\
43 & UZB1 & Maidanak & 10832 & 7078 & 2804 \\
44 & MEX2 & Yucatan & 9046 & 4149 & 530 \\
45 & FR3 & CEA & 1359 & 532 & 13 \\
46 & PER1 & Arequipa & 885 & 584 & 296 \\
47 & BDV & Beidou & 725 & 725 & 725 \\
48 & AUS3 & Carnavon & 450 & 290 & 285 \\
\hline
\end{longtable}

\clearpage
\section{Acknowledgments to the contact persons by site }
\label{app:Acknowledgments}

We would like to express our sincere gratitude to the individuals listed in the following table, who greatly assisted us in deploying and configuring the SVOM stations. Without their invaluable help, the project would not have been able to achieve its objectives. 

\begin{longtable}{p{7cm} p{8cm}}
\caption{List of sites and contact persons} \\
\toprule
\textbf{Sites} & \textbf{Contacts} \\
\midrule
\endfirsthead

\toprule
\textbf{Site} & \textbf{Contacts} \\
\midrule
\endhead

\bottomrule
\endfoot

\bottomrule
\endlastfoot

1. Açores (Portugal, Estação Geodésica Fundamental RAEGE) & Nuno MATA; Nuno Miguel SOARES da MATA \\
\addlinespace

2. Al Aïn (United Arab Emirates, NSSTC) & Saeed ALBLOOSHI \\
\addlinespace

3. Amsterdam (France, TAAF) & Yann LE MEUR; Brendan CORBEL \\
\addlinespace

4. Arequipa (Peru, UNSA) & Pablo Raul YANYACHI \\
\addlinespace

5. Ascension (United Kingdom, ESA, Ariane Station) & Paul BENNETT \\
\addlinespace

6. Athens (Greece, NTUA) & Maria TSAKIRI; Vangelis ZACHARIS \\
\addlinespace

7. Bermuda (ESA) & Frank WILLIAMS \\
\addlinespace

8. Cairns, Darwin, Carnarvon (Australia, UWA) & John MOORE; Bruce GENDRE \\
\addlinespace

9. Chamela, Yucatán, San Pedro Martir (Mexico, UNAM) & Alan WATSON; Fernando ANGELES \\
\addlinespace

10. Chichi-jima (Japon, KU) & Daisuke YONETOKU \\
\addlinespace

11. Djibouti (CERD) & Abalyazid AHMAD; Saad IBRAHIM AHMED; Farah MAHAMOUD OSMAN; Mohamed JALLUDIN \\
\addlinespace

12. Easter Island, Coco Island, Guanhaes, Tristan Da Cunha, Samoa, Marshall Islands (EnviroEarth) & Maxime Le MAILLOT; Maxime GRILLANDINI \\
\addlinespace

13. Diego Garcia (United Kingdom, Sure Limited) & JoJo DELAPLANA; Paul EXALA; Dave HAWORTH \\
\addlinespace

14. Galapagos (Équator, USFQ) & Luis TASIPANTA; Cristina VACA; Leo ZURITA-ARTHOS \\
\addlinespace

15. Hartebeesthoek (South Africa, SANSA) & Raoul HODGES; Carlos DE OLIVEIRA; Tiisetso MASEKO; Frikkie MEYER; Tiaan STRYDOM \\
\addlinespace

16. Hawaii (United State of America, CFHT) & Windell JONES-PALMA; Jean-Gabriel CUBY \\
\addlinespace

17. Ho Chi Minh (Vietnam, VGU) & Hien Vo BICH; Nhu TRAN QUANG \\
\addlinespace

18. Kourou (France, CNES) & Sébastien LACOUR; Julie RICHARD; Romain DELORDRE \\
\addlinespace

19. La Réunion (France, IPGP) & Frédérick PESQUEIRA; Patrice BOISSIER; Christophe BRUNET; Philippe KOWALSKI \\
\addlinespace

20. Le Lamentin (France, Météo France) & Ronan LE MAREC; José ODONNAT; Carole JULIARD \\
\addlinespace

21. Lhassa (China, Tibet University) & Hua BAO; Tianlu CHEN; Shifeng WANG; Jianyan WEI \\
\addlinespace

22. Libreville (Gabon, CNES) & Dominique-Roland DELMAS; Michel DUVERGER; Eric LE GOFF \\
\addlinespace

23. Mahé (Seychelles, SMA) & Vincent AMELIE; Nelson LALANDE; Varunakumar RAJENDRAN \\
\addlinespace

24. Maidanak (Uzbékistan, UBAI) & Samar ABDURAIMOV; Otabek BURKHONOV; Yusufjon TILLAYEV; Dima ALEKSEEV; Bahodir HAFIZOV \\
\addlinespace

25. Malé (Maldives, MMS) & Abdul MUHSIN RAMIZ; Abdulla WAHID; Musa SAYD \\
\addlinespace

26. Malindi (Kenya, BSC) & Munzer JAHJAH; Paolo ROTINI; Davide DECLEMENTE; Boniface OYENGO \\
\addlinespace

27. Manila (Philippines, NAMRIA) & Donnie T. MANCERA; Maria Almalyn A. BALLADARES; Charisma Victoria D. CAYAPAN; Ronaldo GATCHALIAN \\
\addlinespace

28. Marquises (France, Météo France) & Serge MENDIOLA; Fabien CREPIN; William MAUREL \\
\addlinespace

29. Maspalomas (Spain, INTA) & Guillermo RIVERO ARIAS; Maria DE LOS ANGELES DOMINGUEZ DURAN \\
\addlinespace

30. Nanning (China, Guangxi University) & Enwei LIANG; Xiang LU; Lianzhong LÜ \\
\addlinespace

31. Nouméa (France, DITTT) & Perrine DRAIN; Quentin CARDON; Cyrille DUMAS-PILHOU; Thomas MÉNARD \\
\addlinespace

32. Ouagadougou (Burkina Faso, IRD) & Fabrice COURTIN; Mawelamba KATAKA; Boubakar ZEMBA \\
\addlinespace

33. Oukaimeden (Morocco, OUCA) & Zouhair BENKHALDOUN; Omar OUCHAOU \\
\addlinespace

34. Palau (Palau, KARI) & SoonSeob HA; DongHyun KIM; ChunWon KIM \\
\addlinespace

35. Panama (UTP) & Ellis ALICIA; Rodney DELGADO SERRANO; Megan CARRERA \\
\addlinespace

36. Papeete (France, UPF) & Clément MOUNE; Jean-Pierre BARRIOT; Moerani FAILLOUX; Franck MEVEL; Lydie SICHOIX; Youry VERSCHELLE; Yannick VOTA \\
\addlinespace

37. Praia (Cape Verde, NOSI Data Center) & Marta ANDRADE GOMES; Antonio Carlos BARROS LOPES; Luis Carlos CORREIA; Jader SEMEDO \\
\addlinespace

38. Quy Nhon (Vietnam, ICISE) & Son TRAN; Cao SON; Tran NHU; Jean TRAN THENH VAN \\
\addlinespace

39. Rikitea (France, Météo France) & Gilles ORAIN; Tahiariki LEE; Sébastien HUGONY \\
\addlinespace

40. Saint Helena (United Kingdom, MET Office) & Lorimar BENNETT; Marcos HENRY \\
\addlinespace

41. Sharjah (United Arab Emirates, American University of Sharjah) & Georgios CHAMILOTHORIS; Nidhal GUESSOUM \\
\addlinespace

42. Songkhla (Thailand) & Seksan JOMSUREE; Kanthanakorn NOYSENA; Loylip TEERAYUT \\
\addlinespace

43. Tel Aviv (Israel, TAU – Wise Observatory) & Shai KASPI; Dan MAOZ \\
\addlinespace

44. Windhoek (Namibia, H.E.S.S.) & Toni HANKE; Frikkie van GREUNEN \\
\addlinespace

\end{longtable}

\label{lastpage}

\end{document}